\documentclass[showpacs,preprintnumbers,superscriptaddress,amsmath,amssymb]{revtex4-2}
\usepackage[colorlinks=true,bookmarks=false,citecolor=blue,urlcolor=blue]{hyperref}
\usepackage{amsmath,amssymb}
\usepackage{graphicx}
\usepackage{verbatim}
\usepackage{enumitem}
\usepackage{color}
\usepackage[english]{babel}
\usepackage{bm}
\usepackage{natbib}
\usepackage[symbol]{footmisc}
\usepackage{textcomp}

\usepackage{epstopdf}
\usepackage{setspace}
\begin{document}
\title{Super bound states in the continuum through merging in grating}
\author{Evgeny Bulgakov}
\affiliation{Kirensky Institute of Physics, Federal Research Center
KSC SB RAS, 660036, Krasnoyarsk, Russia}
\author{Galina Shadrina}
\affiliation{Institute of Computational Modelling SB RAS,
660036, Krasnoyarsk, Russia}
\author{Almas Sadreev}
\thanks{Corresponding Author}
\email{almas@tnp.krasn.ru} 
\affiliation{Kirensky Institute of Physics, Federal Research Center
KSC SB RAS, 660036, Krasnoyarsk, Russia}
\author{Konstantin Pichugin}
\affiliation{Kirensky Institute of Physics, Federal Research Center
KSC SB RAS, 660036, Krasnoyarsk, Russia}
\date{\today}

\begin{abstract}
Bound states in the continuum (BICs) in gratings composed of
infinitely long silicon rods of rectangular cross-section are
considered. We reveal merging off-$\Gamma$ Friedrich-Wintgen BIC
with symmetry protected BIC. We present CMT and multipole
decomposition theory complementary each other to analyze the
merging phenomenon. The theories show a crossover of the behavior
of $Q$-factor from standard inverse square law $k_{x,z}^{-2}$
towards to extremely fast boosting law $k_{x,z}^{-6}$ in momentum
space. In turn that  crossover gives rise to another crossover
from $Q\sim N^2$ to $Q\sim N^3$ for symmetry protected quasi BIC
in finite grating of $N$ rods owing to suppression of radiation
leakage of quasi BIC mode from surface of grating. As a result the
$Q$-factor of quasi BIC is determined by residual leakage from
ends of grating. We show numerically that this leakage also can be
suppressed considerably if to stretch grating from the ends.
\end{abstract}
\maketitle

\section{Introduction}

Comprehensively tailoring the resonant properties of
electromagnetic resonators are of great importance in fundamental
science and applications \cite{Vahala2003}. The quality ($Q$)
factor of an electromagnetic resonator is a key indicator for
numerous applications. In general, there are several effective
ways to boost the $Q$ factor, for example, whispering gallery
modes in the cavities with convex smooth boundaries such as
cylindrical, spherical or elliptical cavities
\cite{Braginsky1989,Gorodetsky1999}. The another way is to use
Fabry-P\'{e}rot resonator or hide the cavity in photonic crystals
\cite{Vahala2003,Ryckman2012,Seidler2013,Zhou2019}. Cardinally
different way is bound states in the radiation continuum (BICs)
which provide unique opportunity to confine and manipulate
electromagnetic wave within the radiation continuum
\cite{Hsu2016,Azzam2020,Huang2020,Hu2020,Joseph2021,Koshelev2021,Hu2023}.
The phenomenon of BICs is based on that electromagnetic power can
leakage into only selected directions given by diffraction orders
if to arrange dielectric cavities into periodical array
\cite{Hsu2013,Bulgakov2017,Koshelev2019}. Although, the number of
cavities $N$ in the array can not be infinite, $Q$-factor fast
grows with $N$ quadratically for symmetry protected (SP)
quasi-BICs \cite{Taghizadeh2017,Bulgakov2017b,Sadrieva2019} and
cubically for accidental BICs
\cite{Polishchuk,Bulgakov2017b,Sidorenko2021}. However, all these
predictions are breaking down when the non radiative loss
$1/Q_{nr}$ of the photonic crystal (PhC) due to material losses
\cite{Sadrieva2019,Sidorenko2021,Zhang2022} and structural
fluctuations \cite{Ni2017,Maslova2021} surpasses the radiative
loss $1/Q_{r}$ of the system because of $1/Q$ = $1/Q_{nr}$ +
$1/Q_{r}$.  As a result, the non radiative loss will impose a
upper limit of $Q_r$ factor in practice \cite{Sadrieva2019}, which
pinpoints the importance of asymptotic behavior of the $Q$ factor
of BICs over the number of period $N$, i.e. $Q_r(N)\sim
N^{\alpha}$, because improving $Q(N)$ over $Q_{nr}$ does not make
any sense.

\begin{figure}[ht!]
 \centering
\includegraphics[width=0.75\linewidth]{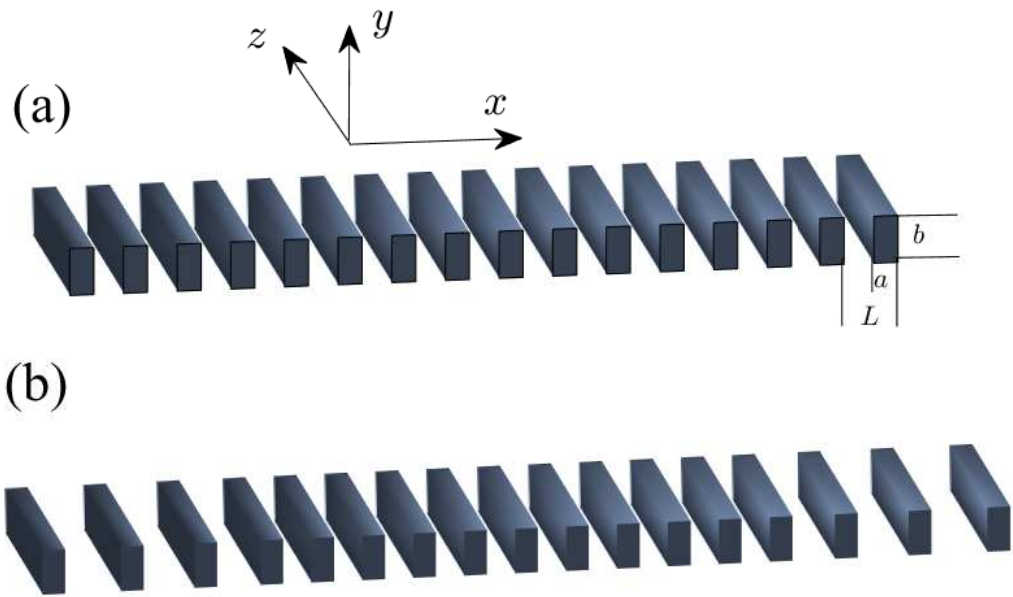}
\caption{Two cases of gratings: (a) with the constant period $L$,
(b) stretched from the ends. The dielectric permittivity of rods
in air $\epsilon=12.11$.} \label{fig1}
\end{figure}

An exploring the ability to boost the Q factor approaching the
upper bound set by the non radiative loss becomes very important.
It is therefore appealing to develop a feasible mechanisms for
enlarging the asymptotic factor $\alpha$. The last time the
phenomenon of merging, at least, two BICs in momentum or
parametric space
\cite{Jin2019,Hwang2021,Kang2021,Bulgakov2022,Huang2022,Zhang2022,Huang2023,
Wang2023,Shubin2023,Barkaoui2023,Fan2023,Zhang2023,Qin2023}
attracted much interest because of crossover of the index $\delta$
in the asymptotic behavior of the $Q$-factor $Q_r\sim
1/(parameter)^{\delta}$ from $\delta=2$ towards $\delta=6$ where
both momentum space or geometrical dimensions of resonators can
serve as a parameter. In turn, merging of BICs forms super BIC
\cite{Jin2019,Koshelev2019a,Zhang2022}. However to the best our
knowledge there were no theory which could show the mechanism of
the crossover in the momentum space for merging BICs. Following to
Ref. \cite{Zhang2022} we present in this paper two alternative
theories complementing each other based on generic two band
effective non Hermitian Hamiltonian (CMT theory) and multipole
decomposition theory with application to grating. In the framework
of the CMT theory we deduce a crossover of the index $\delta$ from
2 towards 6 for approaching merging point that completely agrees
with the results of the multipole decomposition theory.
{Although both theories are generic and can be applied to any
PhC systems which show off-$\Gamma$ BICs due to coupling of two
bands in momentum space we consider in the present paper 1D PhC of
periodical array of dielectric rods sketched in Fig. \ref{fig1}.
Owing to variation of cross-section of rods this PhC system shows
merging off-$\Gamma$ BIC with symmetry protected BIC. In the
framework of the decomposition theory we show that for merging a
full suppression of radiation from surface of grating takes
place.} This result plays a key role for a crossover of asymptotic
behavior of $Q$-factor from $Q\sim N^2$ to $Q\sim N^3$ for finite
grating. Along with that we offer novel mean to suppress also
radiation from the ends by stretching of finite grating from the
ends that considerably boosts further $Q$-factor.
\section{Numerics for avoided crossing of eigenfrequency bands in grating and merging BICs}

\begin{figure}[ht!]
 \centering
\includegraphics[width=0.45\linewidth]{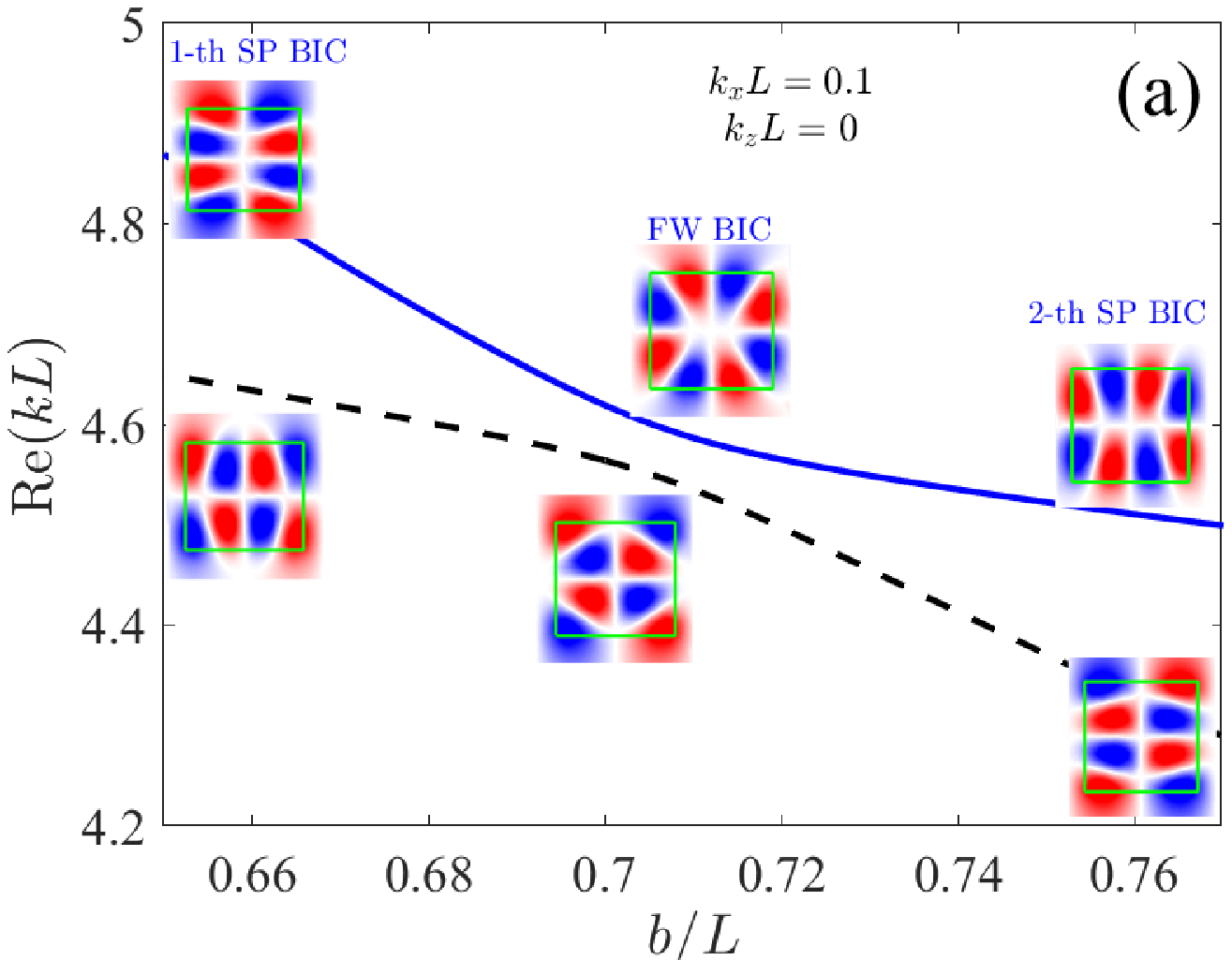}
\includegraphics[width=0.4\linewidth]{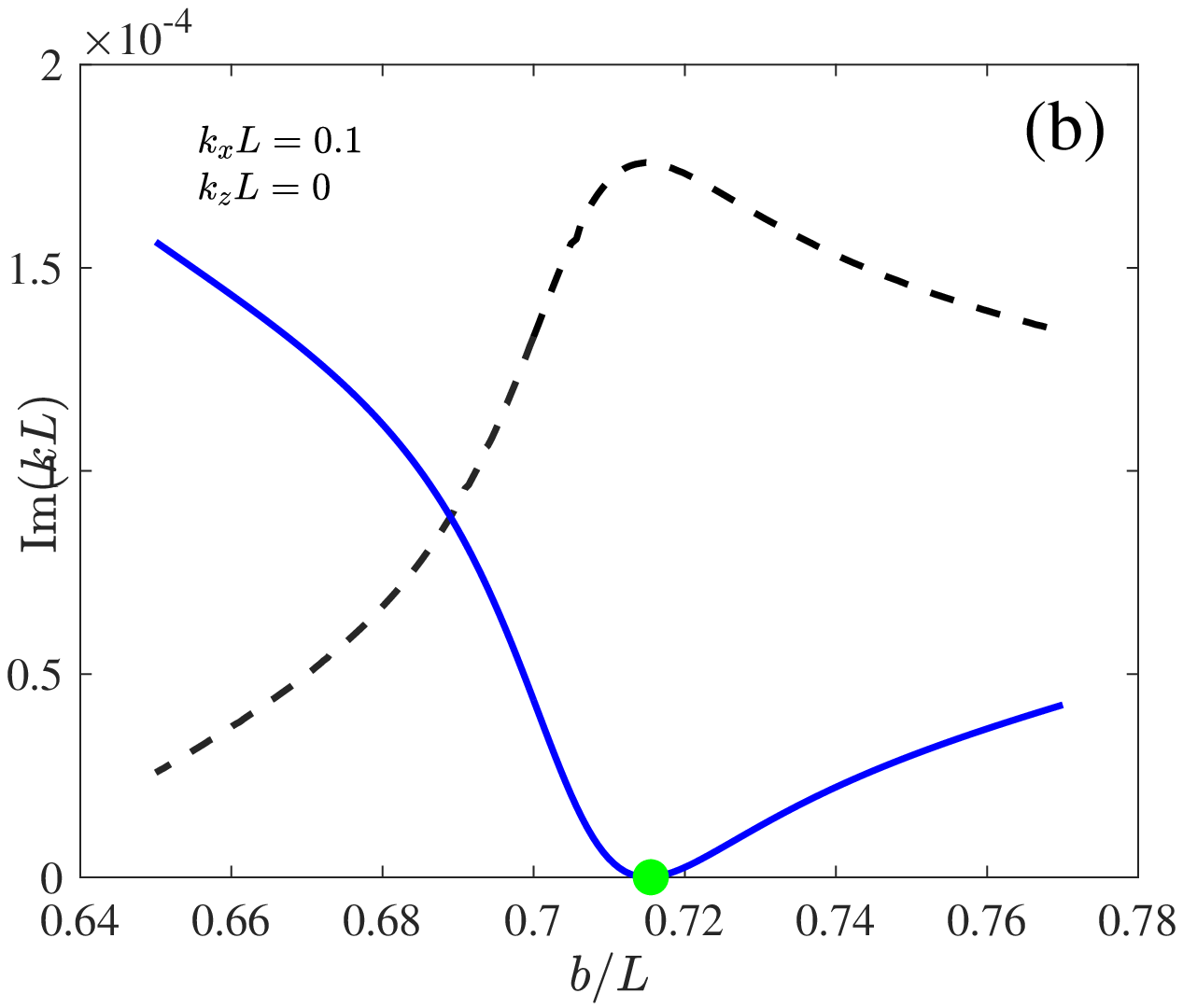}
\includegraphics[width=0.45\linewidth]{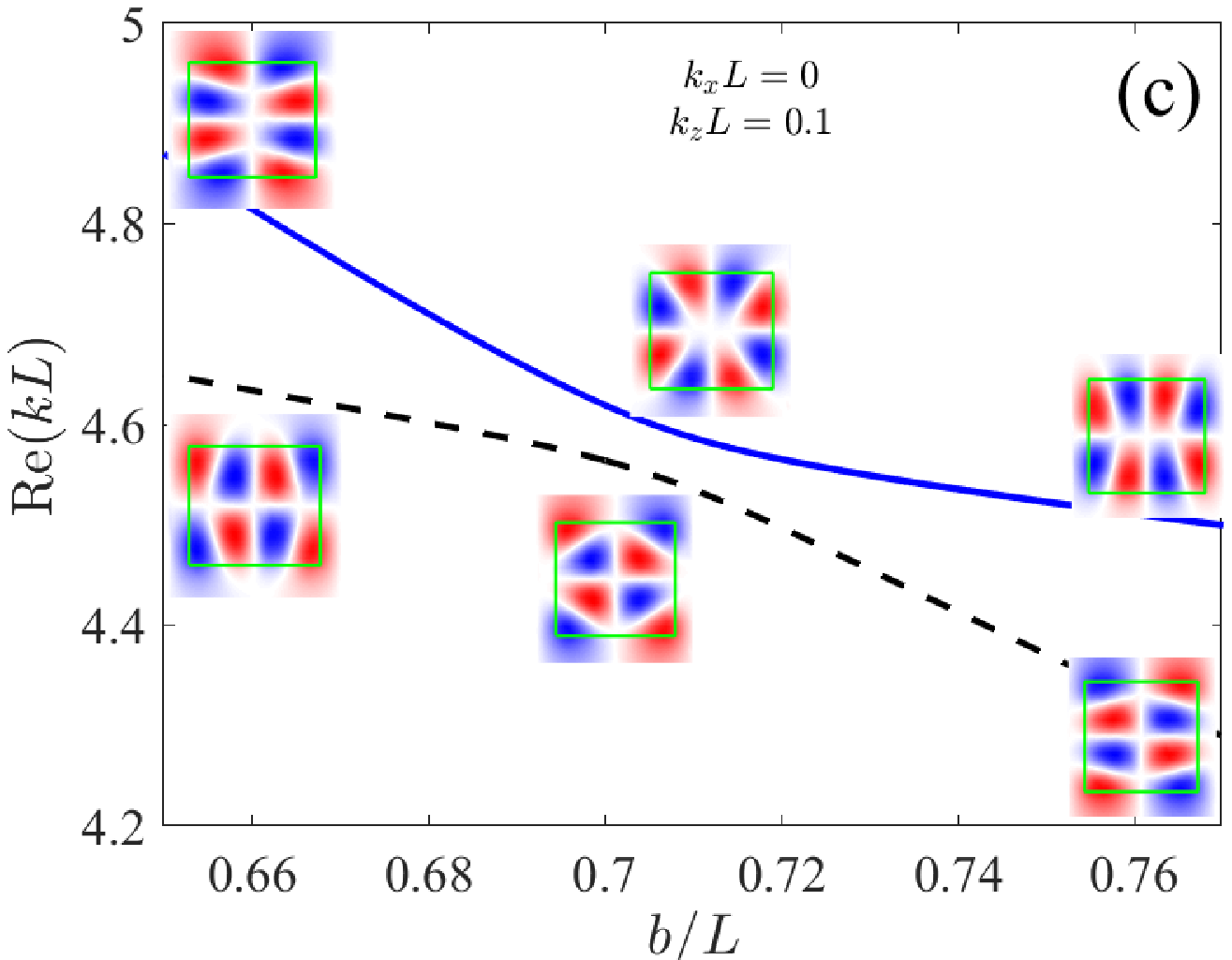}
\includegraphics[width=0.4\linewidth]{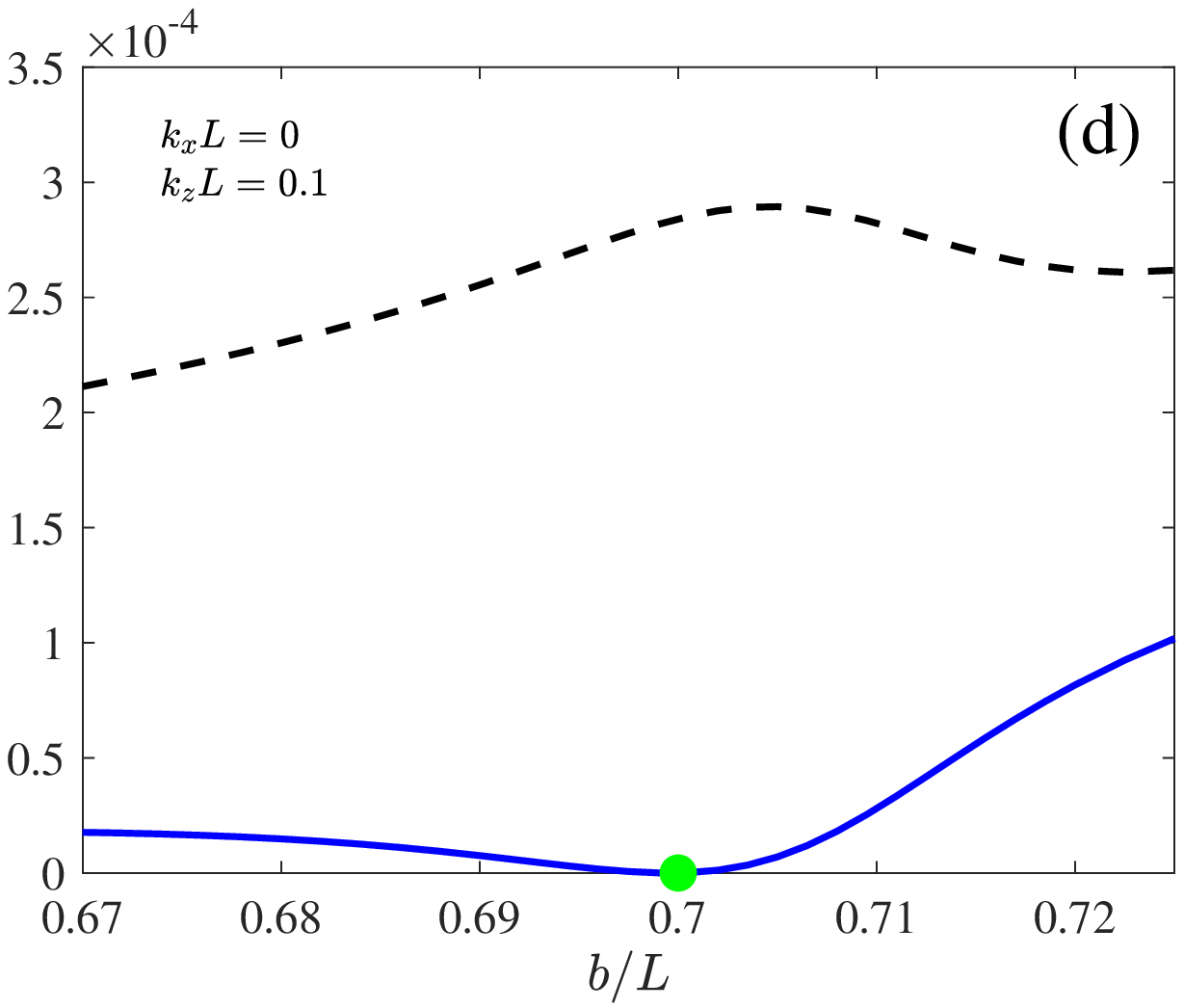}
\caption{ACR of two resonant modes vs height of rod $b$ at
$a/L=0.75$ in the infinite grating. Insets of electric field ${\rm
Re}(E_z)$ illustrate hybridization of resonant modes for ACR.
Green closed circles mark FW BICs.} \label{fig2}
\end{figure}

One of interesting features of open dielectric cavity is that for
variation of its shape the real parts of complex eigenfrequencies,
resonant frequencies undergo ACR accompanied by strong
redistribution of imaginary parts of the complex eigenfrequencies.
As a result the Q-factor can be strongly enhanced
\cite{Wiersig2006,Song2010,Rybin2017,Chen2019,Wang2019,Odit2020,Volkovskaya2020,Huang2021}
forming super cavity modes due to hybridization of resonant modes.
All these features refer also to the present system of array of
rods for variation of the height of rods as demonstrated in Fig.
\ref{fig2}. Insets in Fig. \ref{fig2} (a) and (c) show
respectively hybridization of resonant eigenmodes owing to
interaction through radiation diffraction continua.

\begin{figure}[ht!]
 \centering
\includegraphics[width=0.4\linewidth]{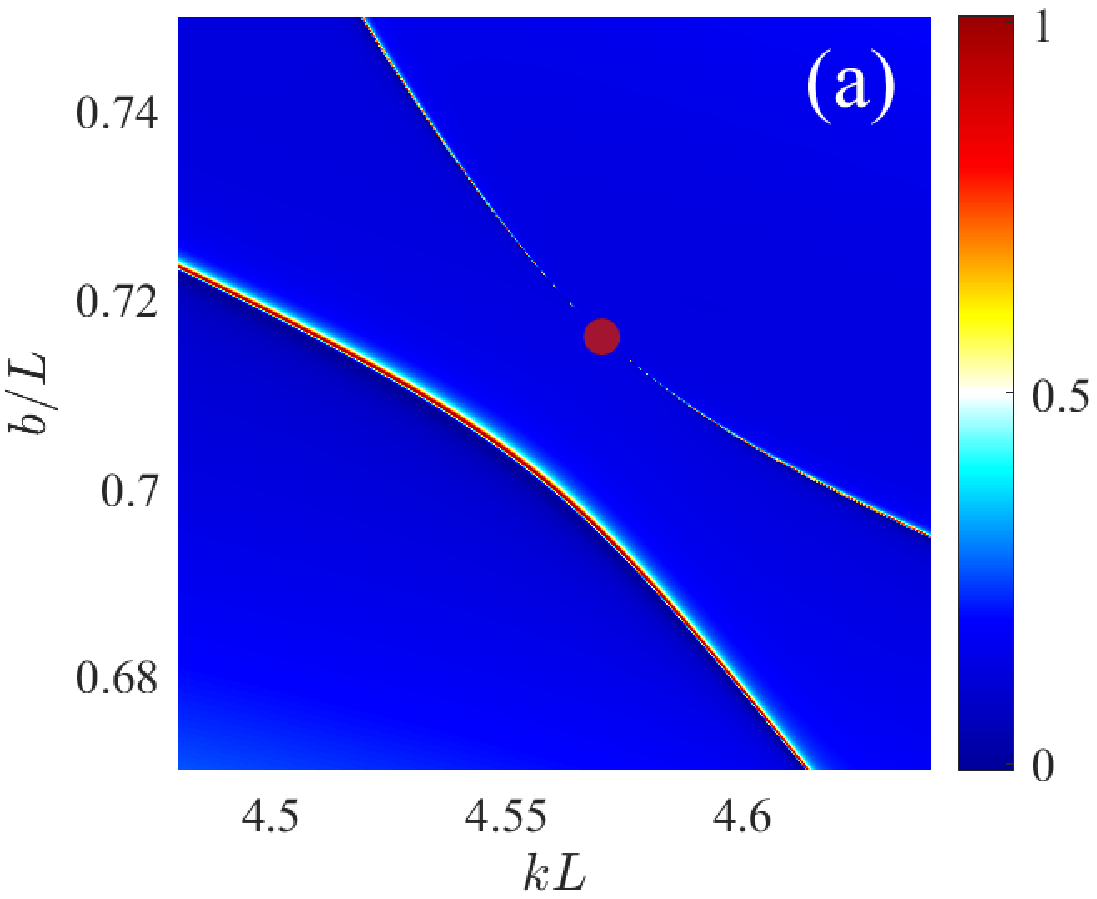}
\includegraphics[width=0.4\linewidth]{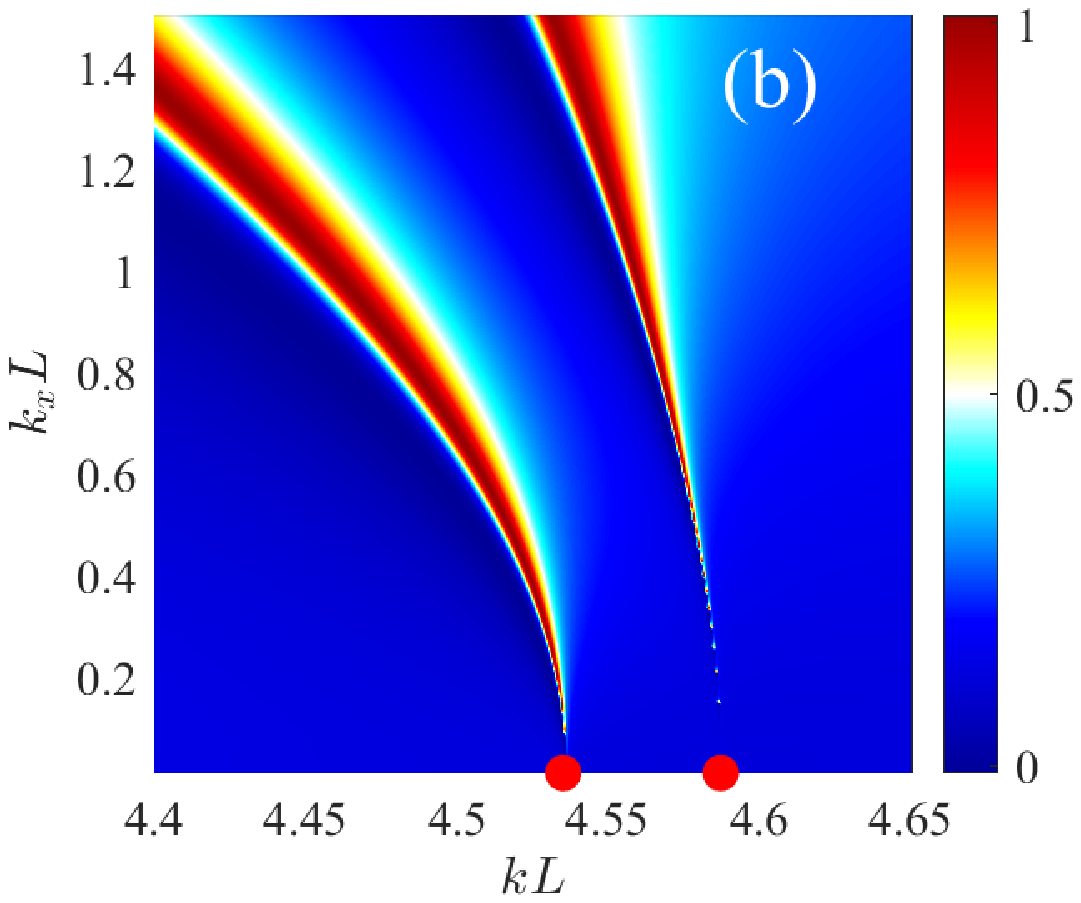}
\caption{(a) Transmittance of TM plane wave  with electric field
directed along $z$-axis vs frequency agreen nd aspect ratio of
rods at $k_xL=0.25, k_z=0$ where circle marks FW BIC. (b)
transmittance vs Bloch vector $k_x$  at $b/L=0.71$ and $k_z=0$.
Closed circles mark SP BICs at $\Gamma$-points.} \label{fig3}
\end{figure}

A grating of infinitely long rods is specified by eigenfrequency
bands which can be clearly seen in transmittance of plane wave
through the grating as Fig. \ref{fig3} shows. Coupling of the
eigenmodes with the radiation continuum leads to ACR of bands that
in turn can give rise to Friedrich-Wintgen (FW) BICs beyond
$\Gamma$-point.
\cite{Hsu2013,Bulgakov2014,Hu&Lu2015,Bulgakov2017,Koshelev2019}.
Moreover, the bands can be featured by symmetry protected BIC at
$\Gamma$-point owing to the symmetry mismatching of the
corresponding eigenmodes with the radiation continuum of the first
diffraction channel
\cite{Vincent1979,Bulgakov2014,Yang2014,Bykov2015,Gao2016,Ni2016}.
All BICs are marked in Fig. \ref{fig3} by closed circles where
evolution of mode profiles is shown in Fig. \ref{fig2} (a) and
(c). Although ACR in Fig. \ref{fig2} is shown beyond
$\Gamma$-point, quite similar ACR takes place at the
$\Gamma$-point. Correspondingly, mode profiles in Fig. \ref{fig2}
(a) and (c) are very close to the true BICs. As Fig. \ref{fig2}
(b) and (d) shows FW BICs occur in both directions of momenta
space.

\begin{figure}[ht!]
 \centering
\includegraphics[width=0.4\linewidth]{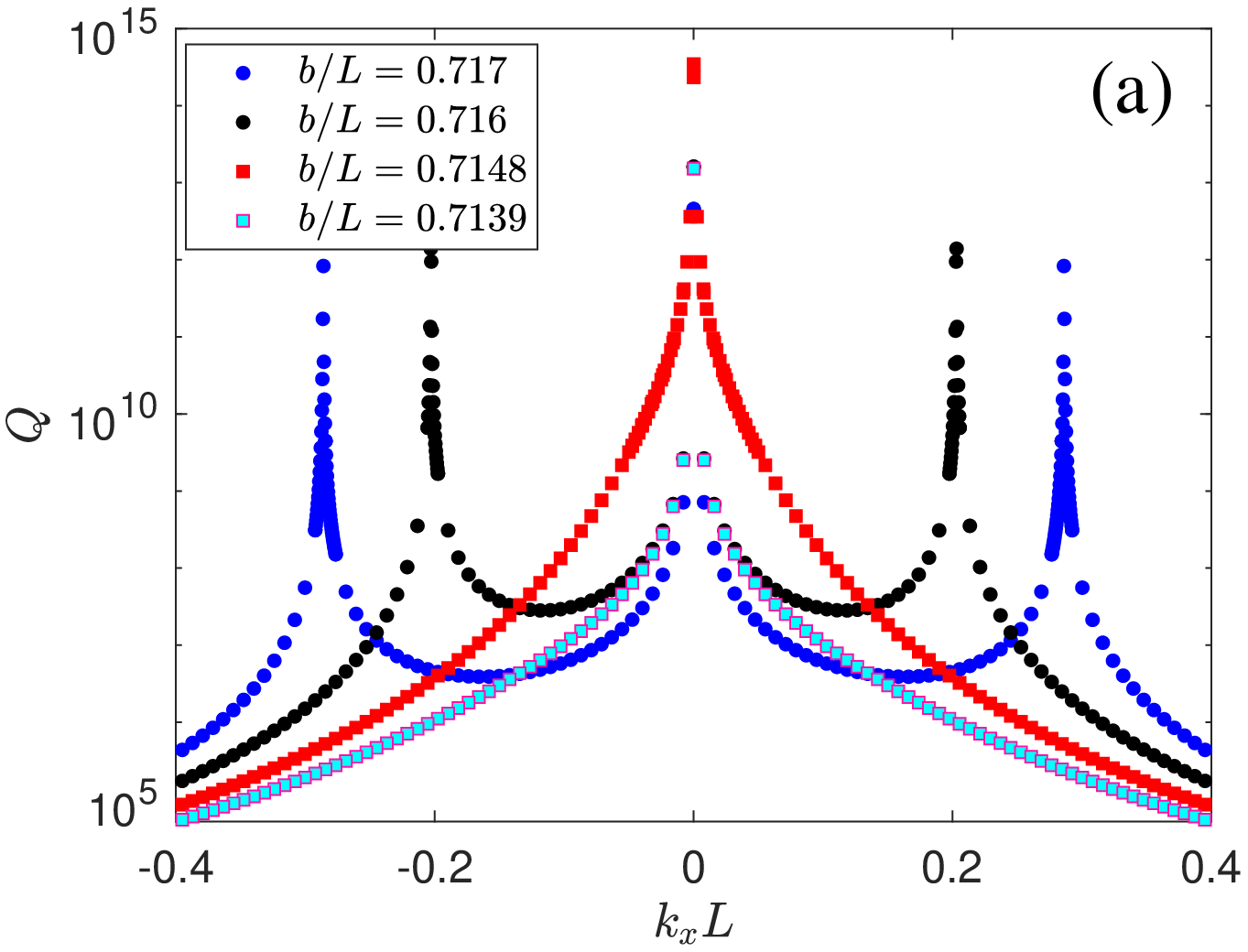}
\includegraphics[width=0.4\linewidth]{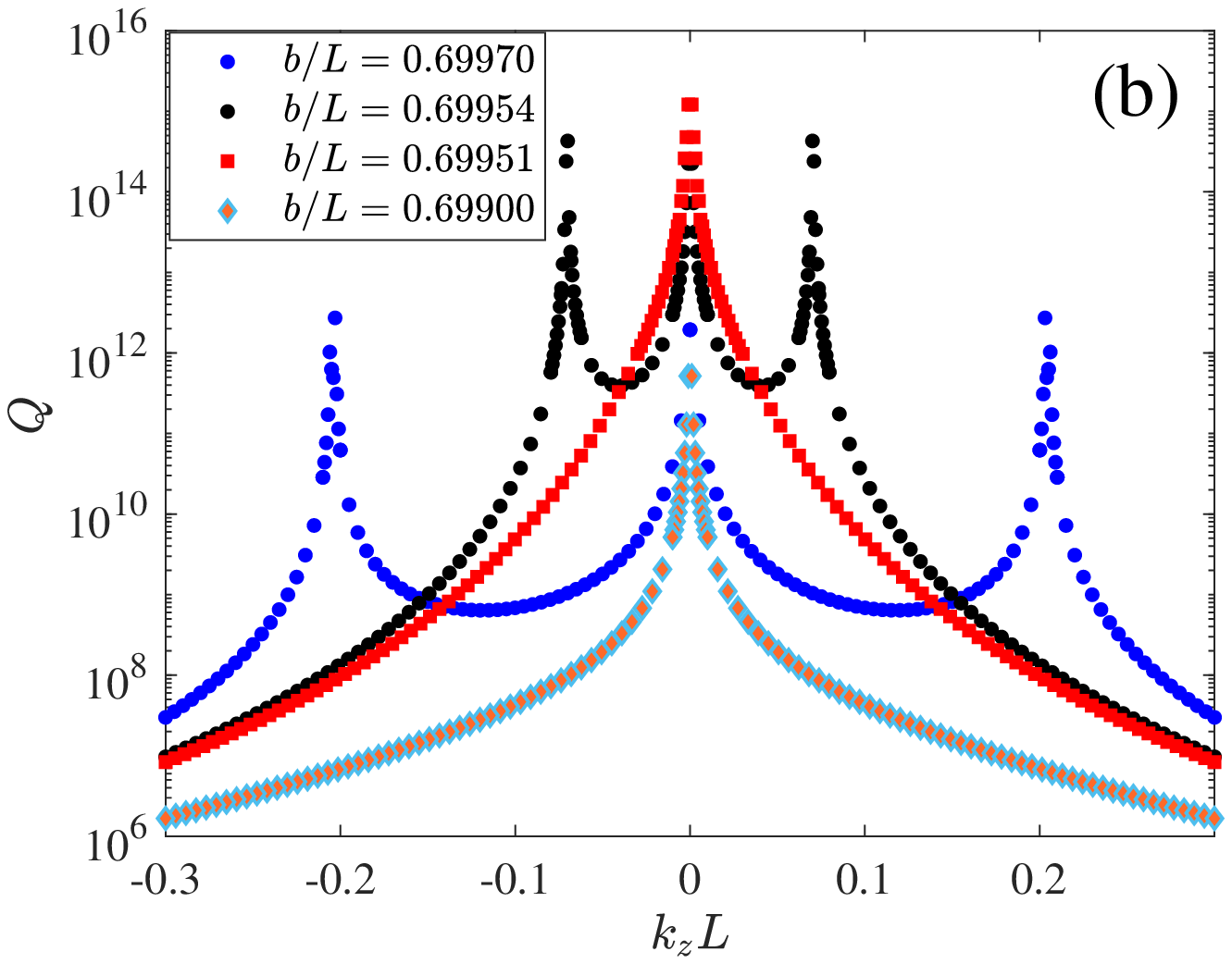}
\caption{Merging FW BIC and SP BIC over Bloch wave number $k_x$ at
$k_z=0$ (a) and wave vector $k_z$ at $k_x=0$ (b).} \label{fig4}
\end{figure}

\begin{figure}[ht!]
 \centering
\includegraphics[width=0.4\linewidth]{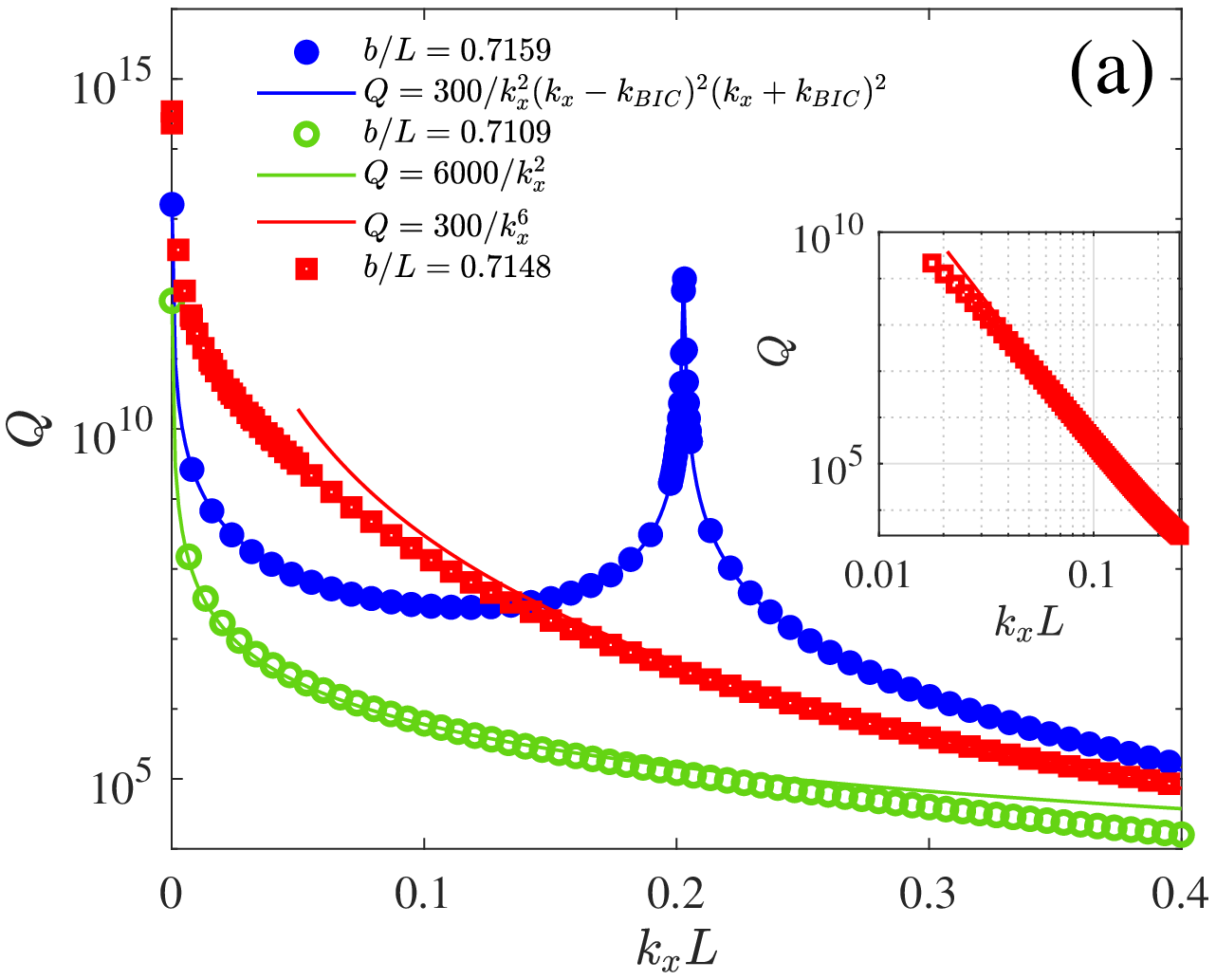}
\includegraphics[width=0.41\linewidth]{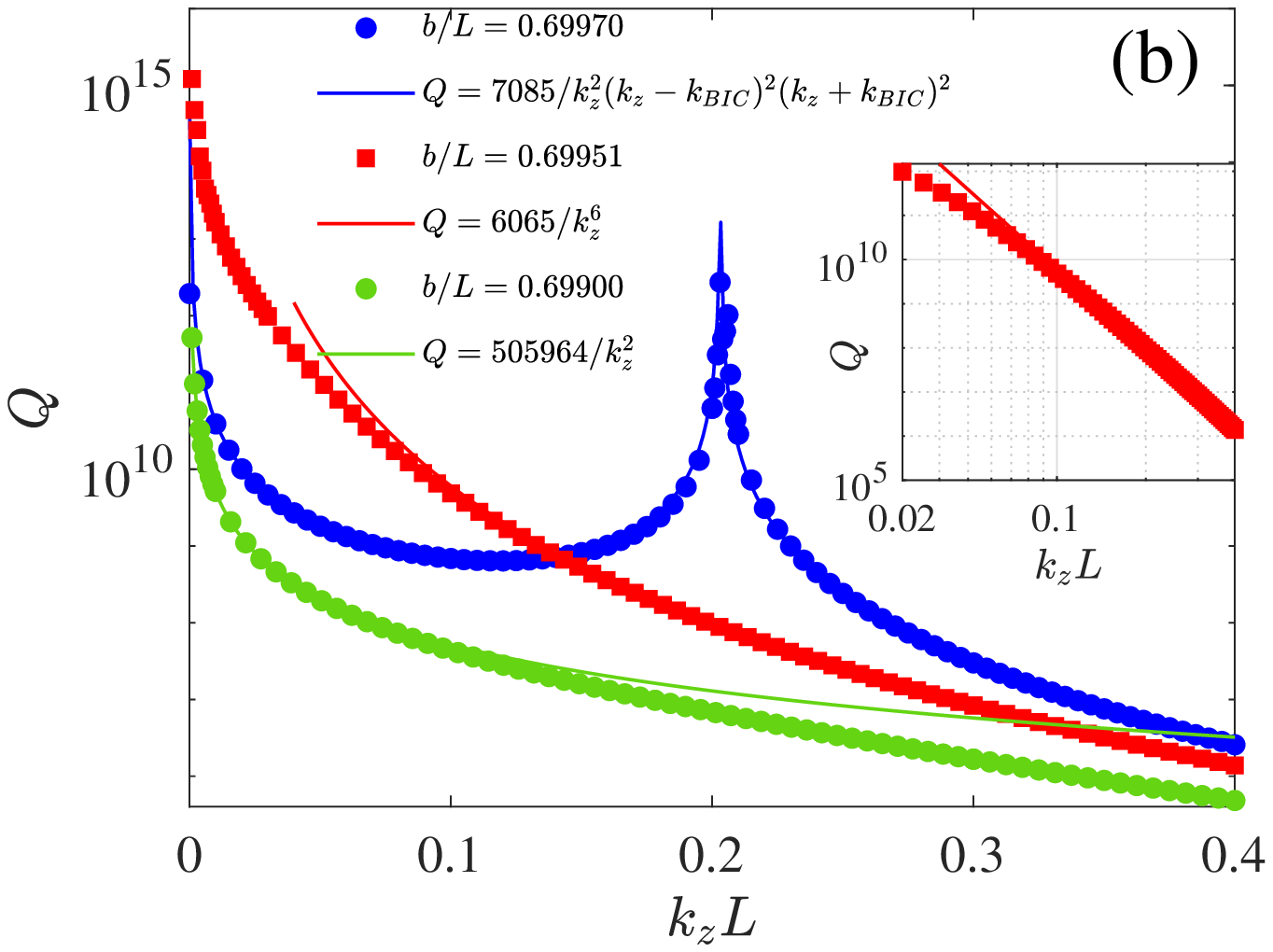}
\caption{Strong redistribution of $Q$-factor on Bloch wave vector
(a) and waveguide vector $k_z$ (b) at merging.} \label{fig5}
\end{figure}

\begin{figure}[ht!]
 \centering
\includegraphics[width=0.45\linewidth]{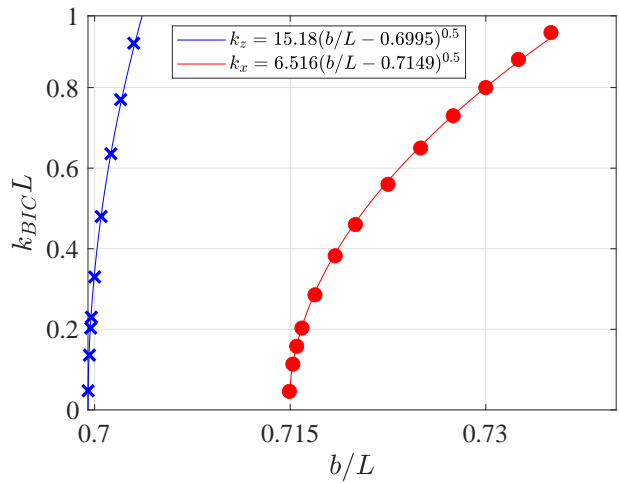}
\caption{The dependencies of momenta of FW BIC on a height $b/L$
of rods at fixed width $a/L=0.75$.} \label{fig6}
\end{figure}

In what follows we focus below on merging of BICs that constitutes
the most interesting and important phenomenon. In Fig. \ref{fig4}
we demonstrate as variation of height $b/L$ of silicon rods the
off-$\Gamma$ FW BIC merges with one of SP BIC at $\Gamma$-point
with $k_x\neq 0$ or $k_z \neq 0$.
One see extremely strong sensitivity of merging to choice of the
ratio $b/L$. Fig. \ref{fig5} illustrates why the phenomenon of
merging is so important because of strong crossover of dependence
of $Q$-factor on wave vectors $k_x$ at $k_z=0$ and $k_z$ at
$k_x=0$ from $Q \sim 1/k_x^2, 1/k_z^2$ to $Q\sim 1/k_x^6, 1/k_z^6$
as insets persuade for limit to the merging points.
Beyond merging point the $Q$-factor can be approximated as
$$Q\sim \frac{1}{k_x^2(k_x-k_{x,BIC})^2(k_x+k_{x,BIC})^2}$$
as was demonstrated by Jicheng Jin {\it et al} numerically in 2d
metasurface \cite{Jin2019}. Below we derive this dependence
analytically based on multipole decomposition theory. Note,
similar dependence of $Q$-factor refers to $k_z$ as Fig.
\ref{fig5} (b) shows. These results for merging FW BIC and SP BIC
are expressed as the dependence of wave vectors on structural
parameter $b/L$ of rods in Fig. \ref{fig6}.

\section{The CMT theory of merging and super BICs}

\begin{figure}[ht!]
 \centering
\includegraphics[width=0.4\linewidth]{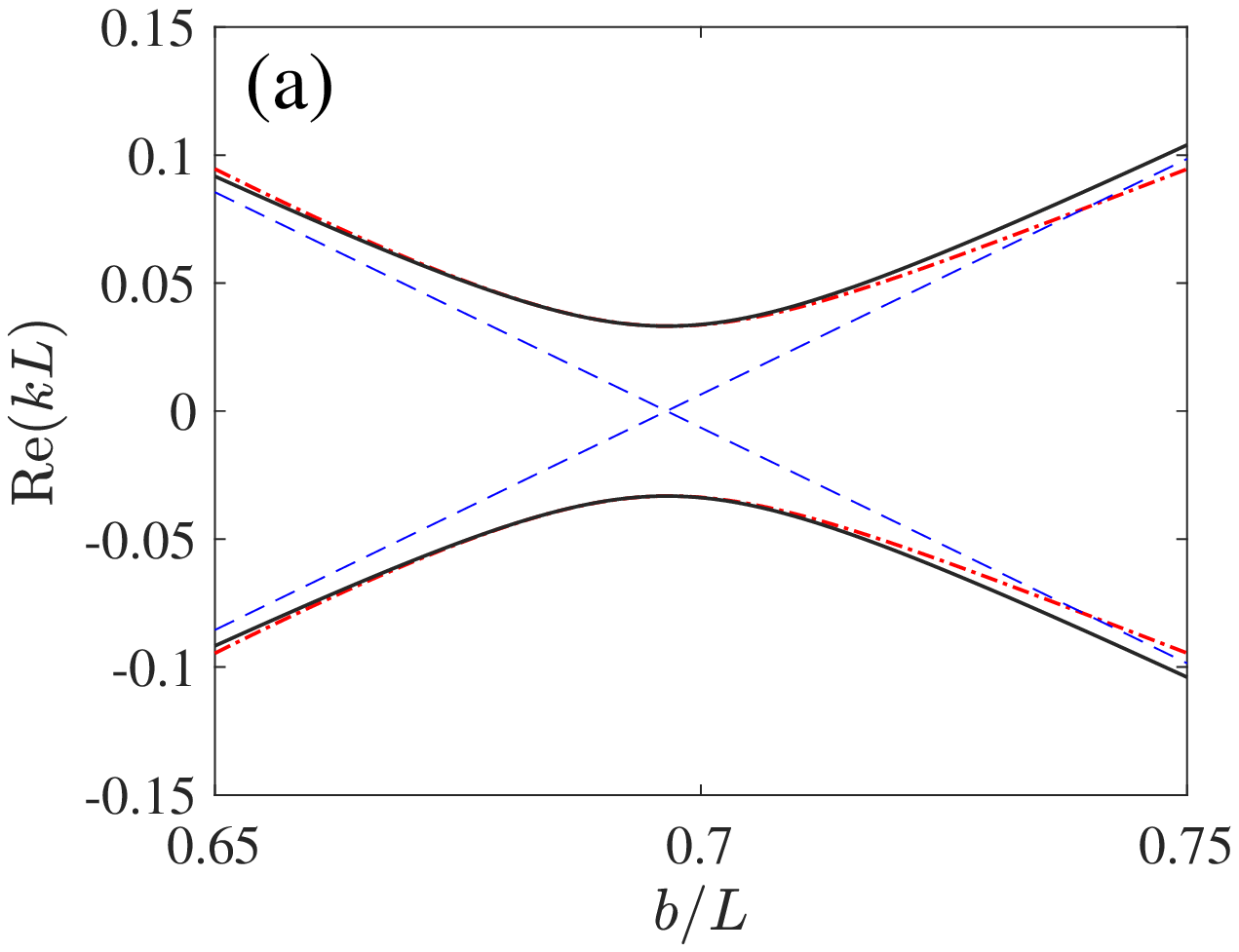}
\includegraphics[width=0.4\linewidth]{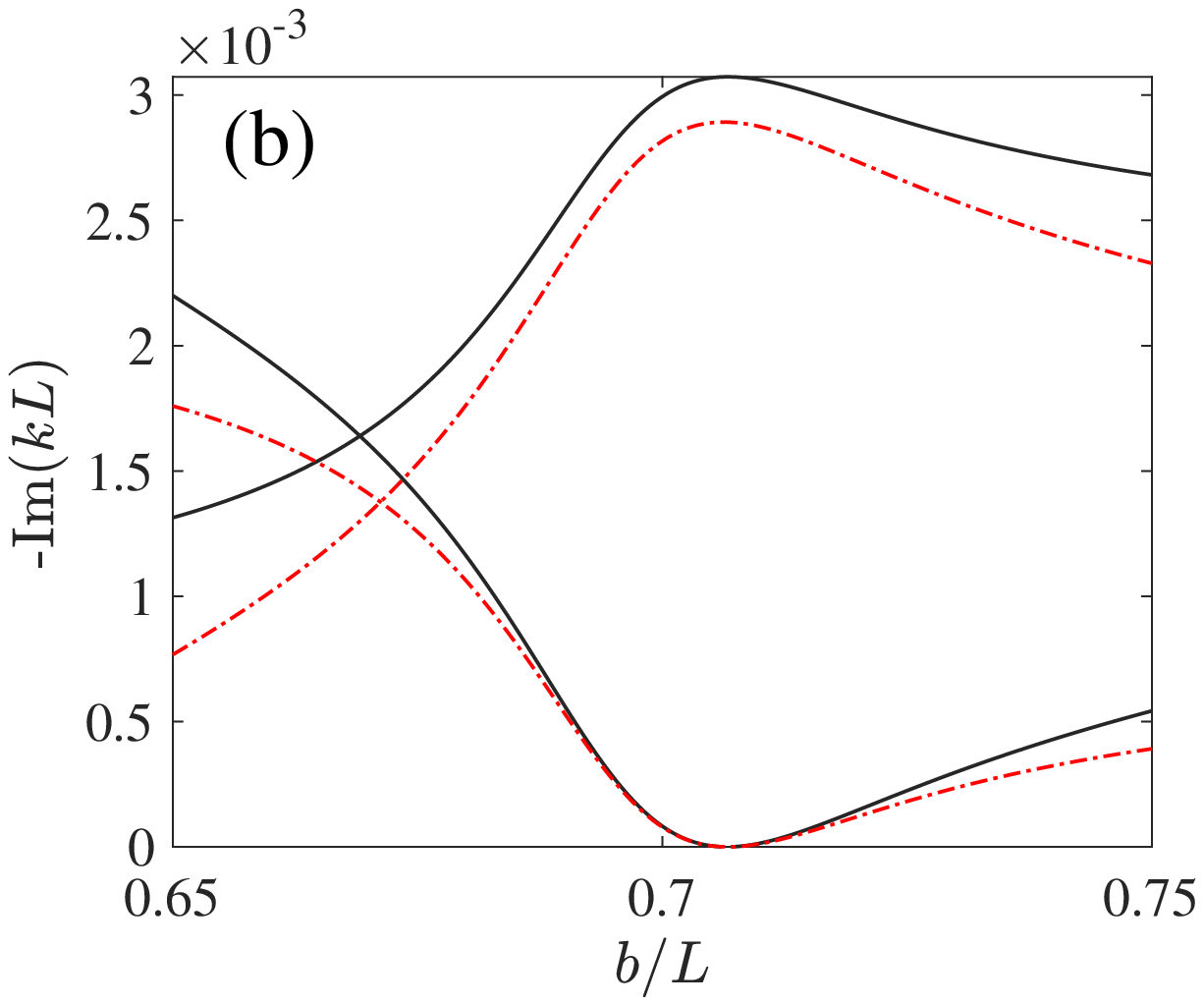}
\caption{Real (a) and imaginary (b) parts of two complex
eigenvalues at $k_xL=0.4, k_z=0$. Solid lines show calculated
numerically at $a/L=0.75$ while dash-dotted lines show fitted
behavior to result in $\varepsilon=-1.918b/L+1.337, e=0.027,
\gamma_1=0.0143, \gamma_2=0.00487, u=0.0336$.} \label{fig7}
\end{figure}

In order to qualitatively describe merging off-$\Gamma$ FW BIC and
SP BIC we introduce generous two-level description of effective
non Hermitian Hamiltonian following Ref. \cite{Zhang2022}
\begin{equation}\label{H2}
     H_{eff}=\left(\begin{array}{cc} \varepsilon+ek_x^2-i\gamma_1k_x^2 &
    u-i\sqrt{\gamma_1\gamma_2}k_x^2\cr
  u-i\sqrt{\gamma_1\gamma_2}k_x^2 &
  -\varepsilon-ek_x^2-i\gamma_2k_x^2 \end{array}\right)+
  \lambda\left(\begin{array}{cc} 1 & 0\cr
 0 & 1 \end{array}\right).
\end{equation}
Here to begin with we put $k_z=0$. The parameters
$\varepsilon(b/L), \lambda(b/L)$ and $e$ response for two PhC
bands at $\Gamma$-point, and $\gamma_{1,2}k_x^2$ describe leakage
of modes for deviation from $\Gamma$-point. Although the SP BICs
at $\Gamma$-point are decoupled from the first radiation continuum
they interact through the next closed diffraction continua that is
expressed by the coupling constant $u$. This Hamiltonian is widely
used for description of FW BICs
\cite{Friedrich1985,Volya,Sadreev2021} however it holds important
novel contribution of dispersive resonant eigenmodes of the
grating. The quantitative values of all model constants in the
Hamiltonian (\ref{H2}) can be extracted from numerically
calculated complex eigenfrequencies of the grating and given  in
Fig. \ref{fig7} where a contribution of the trivial second part of
unit matrix is disregarded.

The complex eigenfrequencies of the effective Hamiltonian
(\ref{H2}) equal
\begin{equation}\label{Z12}
    Z_{1,2}=-i\gamma k_x^2\pm \sqrt{(\varepsilon +ek_x^2-i\delta\gamma k_x^2)^2+
    (u-i\sqrt{\gamma_1\gamma_2}k_x^2)^2}
\end{equation}
describe two resonances whose imaginary parts or resonant widths
versus $\varepsilon$, i.e., aspect ratio $b/L$ and wave vector
$k_x$ are shown in Fig. \ref{fig7}. Here
$\gamma=\frac{\gamma_1+\gamma_2}{2},
\delta\gamma=\frac{\gamma_1-\gamma_2}{2}$.
At $k_x=0$ the model describes two SP BICs for any $\varepsilon$
in the correspondence to Fig. \ref{fig3} (b). Moreover the model
describes also one off-$\Gamma$ BIC of the Friedrich-Wintgen
origin due to avoided crossing of two bands for specific $k_x$
which depends also on $\varepsilon$. That occurs at
\begin{equation}\label{FWBIC}
\varepsilon=\tilde{u}\delta\gamma-ek_{x,BIC}^2,
\end{equation}
where $\tilde{u}=\frac{u}{\sqrt{\gamma_1\gamma_2}}$. This equation
follows from equation for FW BIC derived in Refs.
\cite{Volya,Kikkawa2019,Sadreev2021}. Moreover the Eq.
(\ref{FWBIC}) predicts square dependence of the structural
parameter $\varepsilon(b/L)$ on wave vector of FW BIC that
completely agrees with numerics presented in Fig. \ref{fig6}.

\begin{figure}[ht!]
 \centering
\includegraphics[width=0.45\linewidth]{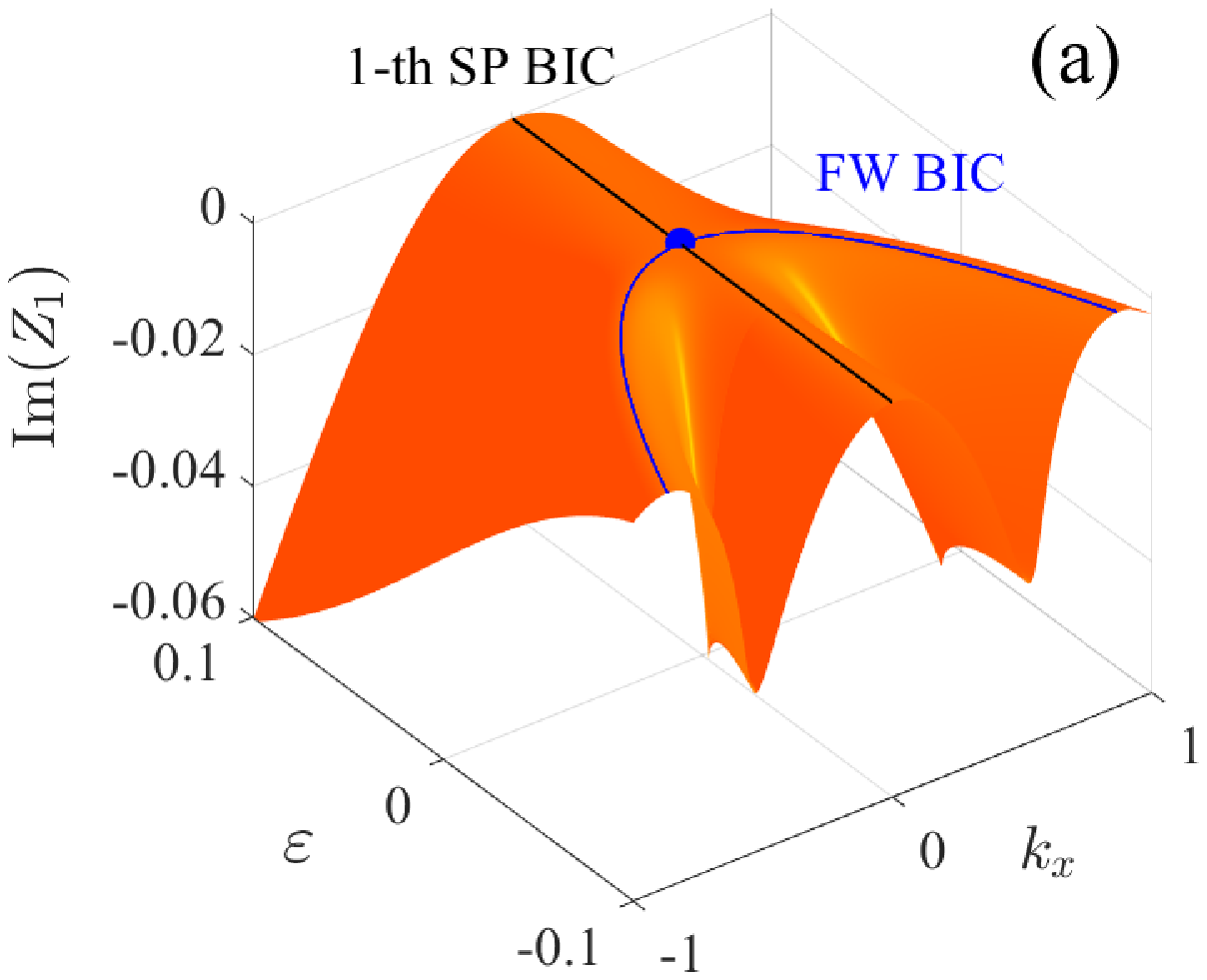}
\includegraphics[width=0.45\linewidth]{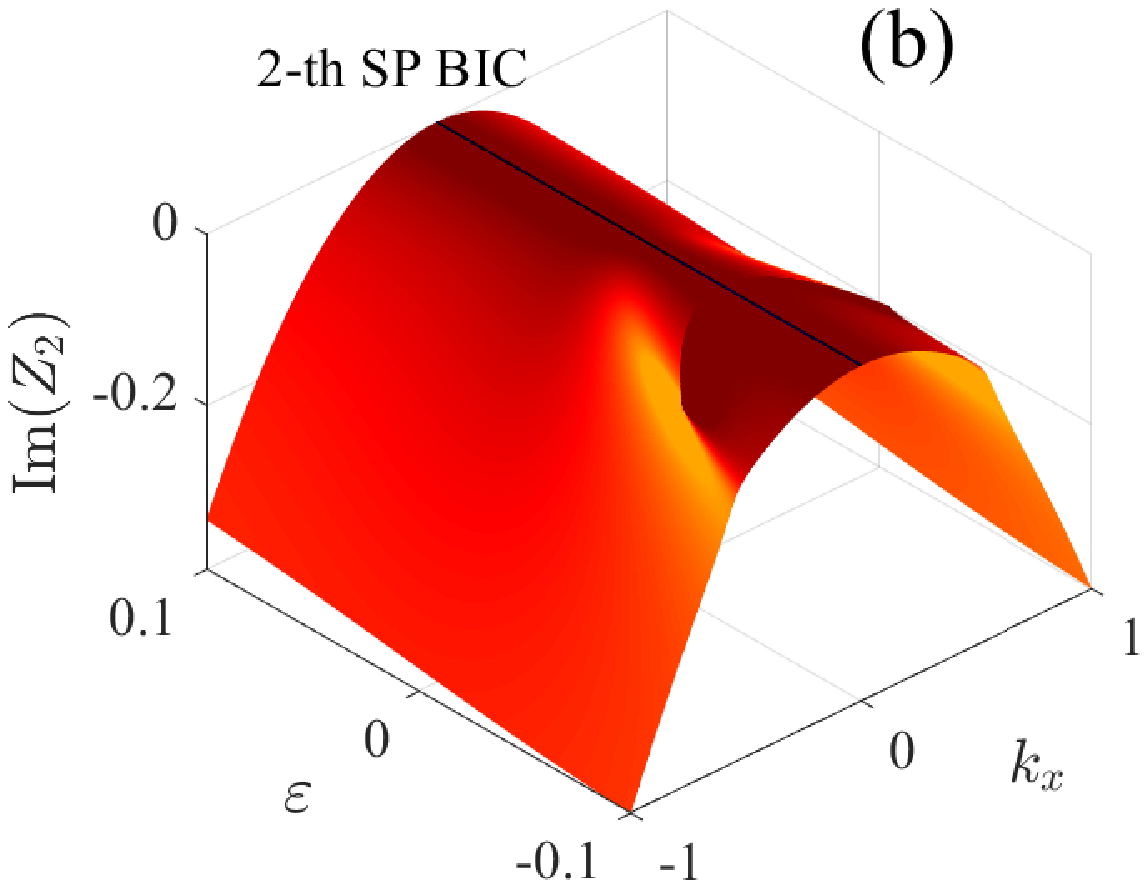}
\caption{(a) and (b) The imaginary parts of two complex
eigenvalues (\ref{Z12}) of the effective Hamiltonian (\ref{H2})
with two SP-BICs and one off-$\Gamma$ FW BIC. Solid lines show SP
and off-$\Gamma$ FW BIC given by Eq. (\ref{FWBIC}). Closed circle
marks merging point. The parameters of Hamiltonian (\ref{H2}) are
chosen as follows: $e=0.15, \gamma_1=0.3, \gamma_2=0.1, u=0.02$.}
\label{fig8}
\end{figure}

However what is the most important, Eq. (\ref{FWBIC}) describes
merging FW BIC with one of SP BIC at $
\varepsilon=\tilde{u}\delta\gamma$ for $k_x\rightarrow 0$ as
illustrated in Fig. \ref{fig8} (a).
Beyond the merging point the imaginary parts of both resonant
modes proportional to $k_x^2$ to give inverse squared behavior
($\delta=2$) of the $Q$-factor as follows from Eq. (\ref{Z12}). At
the merging point the eigenvalues (\ref{Z12}) equal
\begin{equation}\label{mergZ12}
    Z_{1,2}=-i\gamma k_x^2 \pm \gamma(\tilde{u}-ik_x^2)\sqrt{1+\frac{2e\delta\gamma k_x^2}
    {\gamma^2(\tilde{u}-ik_x^2)}
    +\frac{e^2k_x^4}{\gamma^2(\tilde{u}-ik_x^2)^2}}.
\end{equation}
At the vicinity of $\Gamma$-point $k_x\ll 1$ we obtain the
remarkable result of extremely large index $\delta=6$ for resonant
width at the merging point
\begin{equation}\label{kx6}
    Z_1\approx -\gamma\tilde{u}-\frac{\delta\gamma e}{\gamma}k_x^2- i\frac{a^2\gamma_1\gamma_2}
    {2\tilde{u}^2\gamma^3}k_x^6,
\end{equation}
i.e., the $Q$-factor at the merging point grows as $1/k_x^6$.
Thus, the Hamiltonian (\ref{H2}) describes the crossover of the
quality factor $Q\sim 1/k_x^{\delta}$ from $\delta=2$ towards
$\delta=6$. That analytical result agrees with numerical
computation shown in Fig. \ref{fig5} (a) and explains numerical
observations presented for 2d PhCs \cite{Jin2019,Hwang2021}.
Obviously, a similar analytical result can be obtained for
$Q$-factor versus waveguide vector $k_z$ in full agreement with
our numerical computations presented in Fig. \ref{fig5} (b). In
the next section we show that the crossover in suppression of
leakage  at merging BICs plays an important role in the crossover
of asymptotic behavior of $Q$-factor from $N^2$ to $N^3$ for quasi
SP BIC that justifies a terminology of super BIC \cite{Hwang2021}
in grating with finite number $N$ of rods.

\section{Multipole decomposition theory of suppression  of radiation at merging
due to ACR}

The definition of quality factor Q is the ratio of the energy
stored in the system to the power radiation. We consider the
high-refractive index rods and think that the internal energy
stored in the rods is much greater than the external energy stored
outside \cite{Gladyshev2022}. The radiation leakage can be
evaluated via multipole decomposition of scattering function.
\begin{figure}[ht!]
 \centering
\includegraphics[width=0.45\linewidth]{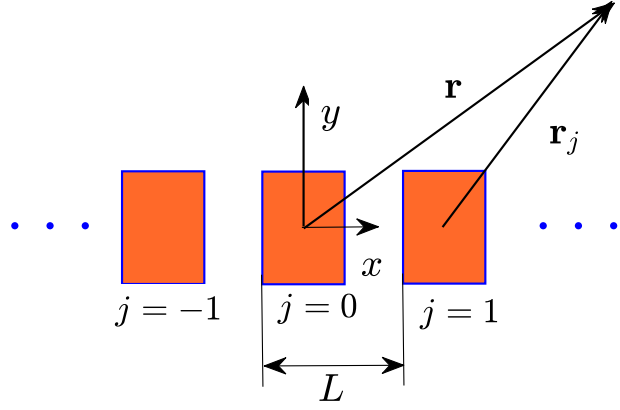}
\caption{Infinite periodical array of rectangular rods.}
\label{fig9}
\end{figure}
\begin{equation}\label{Eq1}
    E_z(x,y)=\sum_j\sum_m a_m
    e^{ijk_xL}H_m(kr_j)e^{im\phi_j},
\end{equation}
where $j$ runs over rectangular rods as sketched in Fig.
\ref{fig9}. Here $r_j$ and $\phi_j$ are the polar coordinates of
the $j$-th radius vector, and ${\bf r}_j={\bf r}-jL{\bf e}_x$.

Using a relationship between the cylindrical harmonic fields and
the space-harmonic fields \cite{Yasumoto} we have for scattering
field
\begin{equation}\label{Eq2}
    E_z(x,y)=\sum_m a_m\frac{2(-i)^m}{Lk^m}\sum_{n=-\infty}^{\infty}\frac{(k_{x,n}+ik_{y,n})^m}
    {k_{y,n}}e^{ik_{x,n}x+ik_{y,n}y}, y>0
\end{equation}
where
\begin{equation}\label{Eq3}
    k_{x,n}=k_x+\frac{2\pi n}{L}, k_{y,n}=\sqrt{k^2-k_{x,n}^2},
  \end{equation}
and integers $  n=0, \pm 1, \pm 2, \ldots$ enumerate diffraction
orders, i.e., radiation continua. In what follows we consider SP
BICs embedded into the first continuum $n=0$ with the
eigenfrequency of BICs $k < 2\pi/L$. The scattering field
(\ref{Eq2}) in the far zone can be  approximated as
\begin{equation}\label{Eq4}
    E_z(x,y)\approx \frac{4}{Lk\cos\theta}\sum_m a_m(k_x)
    e^{-im\theta}e^{ik_xx+ik_yy}=Fe^{ik_xx+ik_yy}, y>0
\end{equation}
where $k_x=k\sin\theta, k_y=k\cos\theta$. Since the scattering
function (\ref{Eq2}) is odd relative to $x\rightarrow -x$ we have
$a_{2m}(k_x)=-a_{-2m}(k_x), a_{2m+1}(k_x)=a_{-2m-1}(k_x)$.
Moreover, $a_{2m}(k_x)=a_{2m}(-k_x),
a_{2m+1}(k_x)=-a_{2m+1}(-k_x)$. Respectively, we have from
(\ref{Eq4})
\begin{equation}\label{Eq5}
  F=\frac{4}{Lk\cos\theta}[-i\sum_{m=1}^{\infty}a_{2m}(k_x)\sin(2m\theta)+
    \sum_{m=0}^{\infty}a_{2m+1}(k_x)\cos((2m+1)\theta)].
\end{equation}

\begin{figure}[ht!]
 \centering
\includegraphics[width=0.45\linewidth]{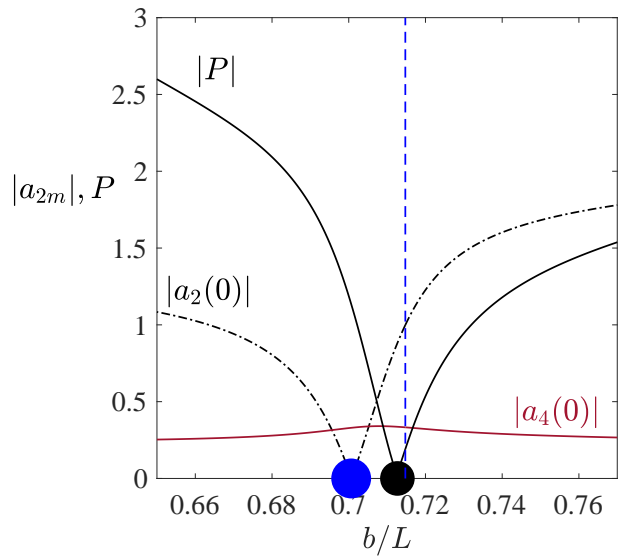}
\caption{Behavior of the first two decomposition coefficients
$a_2(0), a_4(0)$ and the magnitude $P$ in Eq. (\ref{P1}) vs aspect
ratio of rods $b/L$. Dash line corresponds to merging point.}
\label{fig10}
\end{figure}

For slight deviation from the merging point, i.e., for small
$\theta\approx k_x/k$ we obtain from (\ref{Eq5})
\begin{equation}\label{Eq6}
    F\approx P_1k_x+P_3k_x^3,
\end{equation}
where
\begin{equation}\label{P1}
    P_1=-\frac{4i}{k^2L}[\sum_{m=1}^{\infty}2ma_{2m}(0)+
    ik\sum_{m=0}^{\infty}\frac{d a_{2m+1}(0)}{dk_x}]=
    -\frac{4i}{k^2L}P.
\end{equation}
Since the $Q$-factor is a ratio of stored energy $U$ and leaking
power $W=|F|^2$ we have
\begin{equation}\label{Q}
    Q=\frac{kU}{|F|^2}=\frac{kU}{|P_1k_x+P_3k_x^3|^2}.
\end{equation}
The decomposition coefficients $a_m$ can be  expressed via
integral over cross-section of rods \cite{Johnson2001}
\begin{equation}\label{am}
    a_m=\frac{i\pi k^2}{2}\int d\Omega
    J_m(kr)\frac{e^{-im\phi}}{\sqrt{2\pi}}(\epsilon({\bf
    x})-1)E_z^{*}({\bf x})d\Omega.
\end{equation}
The first two coefficients are shown in Fig. \ref{fig10} at
$\Gamma$-point versus $b/L$ from where one can see that $a_2$
undergoes critical behavior owing to ACR.
According to this Figure we can present the coefficients in Eq.
(\ref{Q}) as
\begin{equation}\label{PP1}
P_1=-\beta(b-b_c), P_3=C,
\end{equation}
where $b$ is the structural parameter shown in Fig. \ref{fig1}. As
a result we obtain
\begin{equation}\label{Qperp}
    Q\sim \frac{1}{|-\beta(b-b_c)k_x+Ck_x^3|^2}.
\end{equation}
Thus, for infinite grating we obtain the equation for BIC
\begin{equation}\label{parabol}
    Ck_x^2=\beta(b-b_C)
\end{equation}
at the merging point $b=b_c, k_x=0$.  Moreover for the limit to
the merging point we have
\begin{equation}\label{law}
Q\sim \frac{1}{k_x^6}.
\end{equation}
It is remarkable, from Eq. (\ref{Qperp}) we have
\begin{equation}\label{Jin}
    Q\sim \frac{1}{k_x^2|-\beta(b-b_c)+Ck_x^2|^2}=
    \frac{1}{\beta k_x^2(k_x+k_{BIC})^2(k_x-k_{BIC})^2},
\end{equation}
where $k_{BIC}=\sqrt{\beta(b-b_c)/C}$ that fully agrees with
numerical derivations presented in Fig. \ref{fig5} as well as with
numerically derived expressions by Jicheng Jin {\it et al}
\cite{Jin2019} for 2D PhC.

\begin{figure}[ht!]
 \centering
\includegraphics[width=0.45\linewidth]{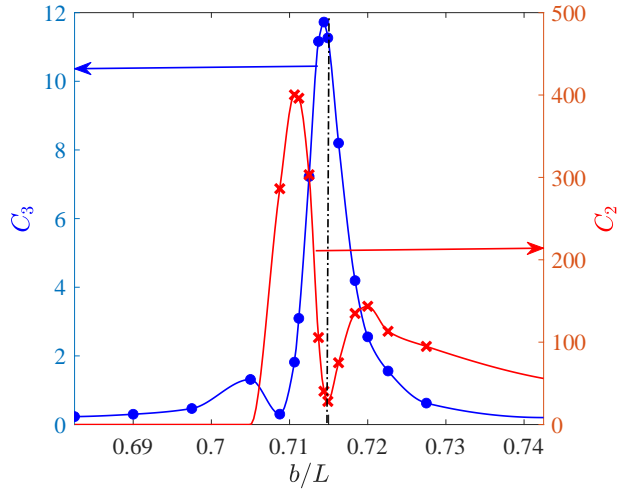}
\caption{Behavior of coefficients $C_2$ (right) and $C_3$ (left)
in dependence of the $Q$-factor $Q=C_2N^2+C_3N^3$ vs the number of
rods in grating $N$. Closed circles and crosses show Comsol
calculations, solid lines show interpolation at the interval of
$N=10$ till $N=100$. Dash line corresponds to merging point.}
\label{fig11}
\end{figure}

Now we consider grating with finite number of rods $N$ and argue
that a change of the index $\delta$ in asymptotical behavior of
the $Q\sim \frac{1}{k_x^{\delta}}$ for merging  BICs results in
the change of the behavior of $Q$-factor over the number of
resonators from quadratic to cubic. We assume that the EM power
radiates from surface of finite grating which has the same origin
as leakage calculated above and from the ends of grating, so
respectively we have for the quality factor \cite{Taghizadeh2017}
\begin{equation}\label{QTag}
    \frac{1}{Q}=\frac{1}{Q_{\perp}}+\frac{1}{Q_{\parallel}}.
\end{equation}
Here the first contribution $Q_{\perp}$ is the contribution of
quasi SP BIC which is a standing wave with the wave number
$k_x=\pi/NL$ \cite{Taghizadeh2017,Sadrieva2019}.
Therefore aside radiation from the surface of finite grating gives
us
\begin{equation}\label{Qper}
   \frac{1}{Q_{\perp}}\sim \frac{D_2}{N^2}+\frac{D_6}{N^6},
\end{equation}
according to Eq. (\ref{Q})in correspondence to  the above
derivations of crossover at a vicinity of merging. At the merging
point the first contribution vanishes to become negligible small
compared to radiation from the ends of finite grating to write
$D_2\sim |b-b_c|$. As it was derived by many scholars
$Q_{\parallel}\sim N^3$ by use of  the tight-binding approximation
\cite{Blaustein2007,Polishchuk,Asenjo2017,Bulgakov2019a}. The
crossover can be traced in numerics by fitting $Q=C_2N^2+C_3N^3$
in the interval for $N$ from 10 till 100 that Fig. \ref{fig11}
illustrates.

\begin{figure}[ht!]
 \centering
\includegraphics[width=0.45\linewidth]{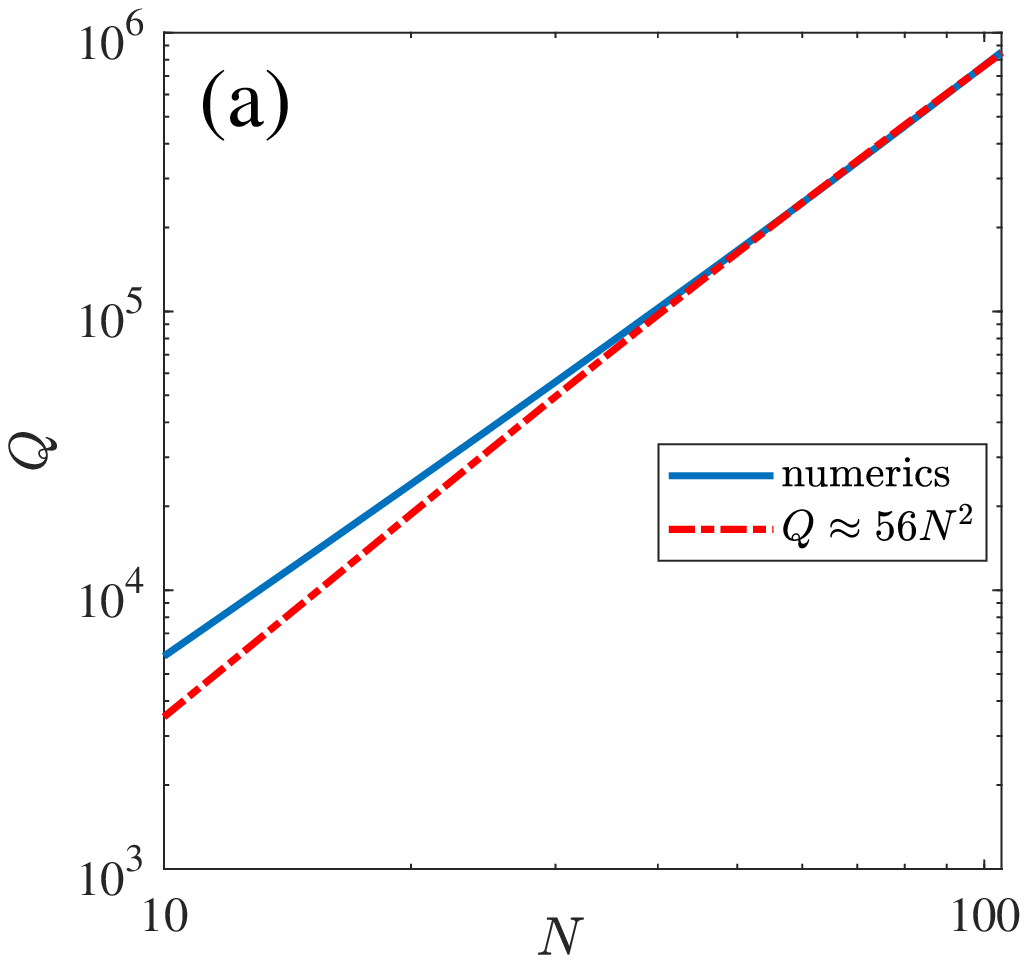}
\includegraphics[width=0.45\linewidth]{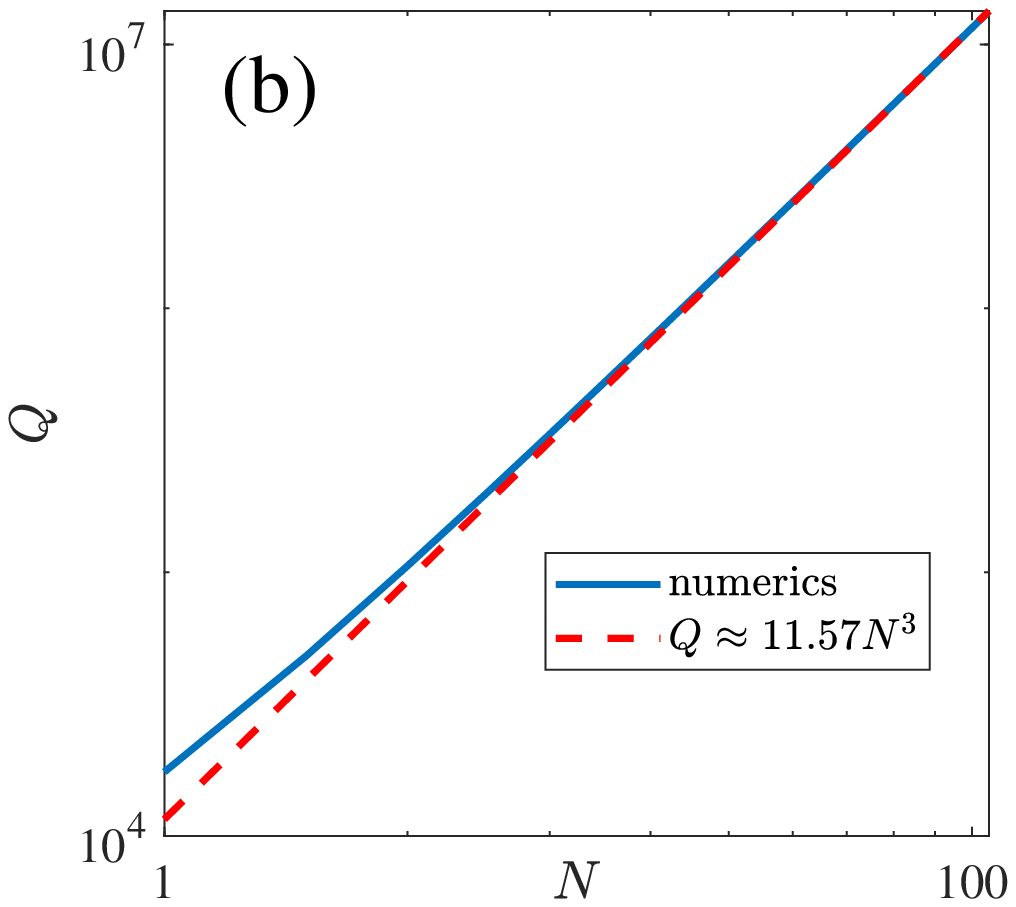}
\caption{Dependence of the Q factor on the number of rods in
finite grating. (a) Far from merging point at $b=0.7425L$ and (b)
at merging $b/L=b_c/L=0.7148$, $a/L=0.75$.} \label{fig12}
\end{figure}

Therefore at merging SP BIC and FW off-$\Gamma$ BIC we obtain
crossover for $Q$-factor from $Q\sim N^2$ to $Q\sim N^3$ resulting
in super quasi BIC as Comsol Multiphysics calculations illustrate
in Fig \ref{fig12}.

\begin{figure}[ht!]
 \centering
\includegraphics[width=0.7\linewidth]{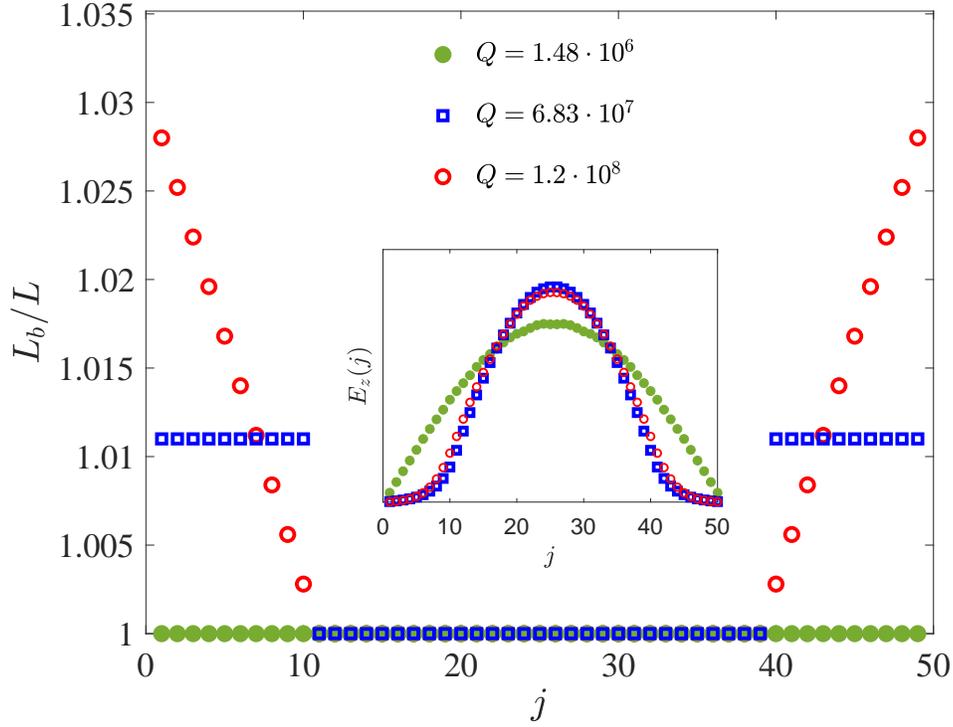}
\caption{The periods between rods $L_b(j)$ of buffer gratings in
terms of the constant period of inner grating $L$. Inset shows
maximal values of mode amplitudes $E_z(j)=max(|E_z(x,y)|)$ inside
$j$-th rod.} \label{fig13}
\end{figure}

\begin{figure}[ht!]
 \centering
\includegraphics[width=1\linewidth]{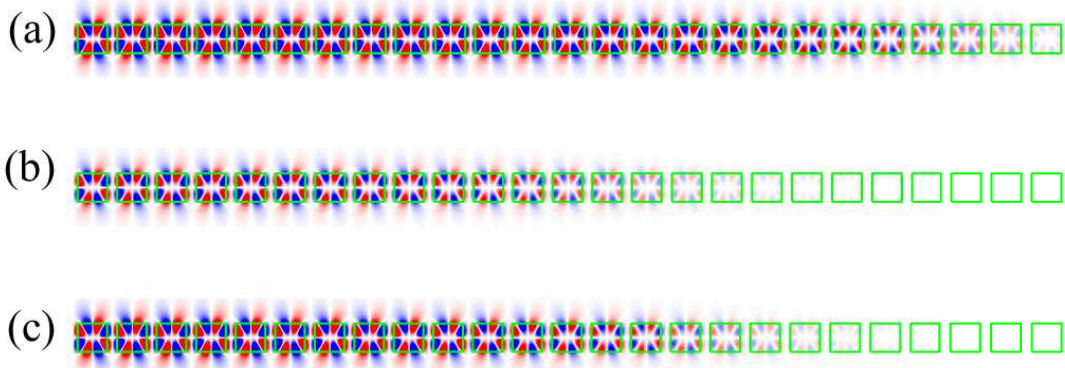}
\caption{(a) The profiles of the solutions (Re[$E_z(x,y)$]) in
finite gratings with the constant period $L$, the grating is
shielded by two gratings with the period $L_b/L=1.015$ (b) as
shown by squares in Fig. \ref{fig13}, and the grating is gradually
stretched from the ends (c) as shown by open circles in Fig.
\ref{fig13}. Because of symmetry only the half of mode profiles
are shown.} \label{fig14}
\end{figure}

\begin{figure}[ht!]
 \centering
\includegraphics[width=0.45\linewidth]{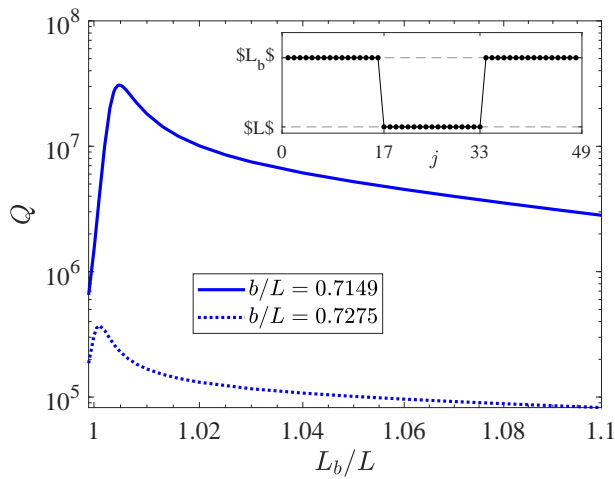}
\caption{$Q$-factor of buffered grating vs period of buffers.
Solid line corresponds to the merging point while dotted line does
beyond merging pont.} \label{fig15}
\end{figure}

Moreover we use additional way to considerably boost the
$Q$-factor by adjustment of additional buffer gratings to the ends
of grating \cite{Zhang2011,Asenjo2017}. These buffer grating have
either the period $L_b$ slightly different from the period $L$ of
basic grating or the period of buffer gratings gradually
stretching as sketched in Fig. \ref{fig1} (b) and shown in Fig.
\ref{fig13}.
That gives rise to strong suppression of the wave function near
the ends of grating as shown in Fig. \ref{fig13} and Fig.
\ref{fig14}. Moreover Fig. \ref{fig13} demonstrates crucial
enhancement of $Q$-factor owing to stretching at merging point
caused by suppression of radiation from ends of grating.
One can see that in spite of very small stretching of grating we
observe strong suppression of wave function at the ends of finite
gratings. As a result we have strong boosting of the $Q$-factor
for increasing of the period of grating of only by  $1 \%$ at
merging as Comsol Multi Physics simulations of $Q$-factor show in
Fig. \ref{fig15}. One can see also from this Figure that
$Q$-factor is boosting much stronger at the merging because of
suppression of surface radiation while radiation from the ends is
suppressed by stretching.

\section{Summary}
We developed a concept of super BIC
\cite{Jin2019,Koshelev2019a,Kang2021} as a result of merging
"usual" Friedrich-Wintgen off-$\Gamma$ BIC with SP BIC at
$\Gamma$-point in the case of grating constituted of silicon rods
of rectangular cross-section. The merging phenomenon reported
recently in application to different systems
\cite{Hwang2021,Kang2021,Zhang2022, Huang2022,Huang2023,
Wang2023,Barkaoui2023,Fan2023,Zhang2023,Qin2023} has attracted
much interest because of crossover of asymptotic behavior of
$Q$-factor over critical parameter. Wave vector which defines
frequency bands in PhCs attracts particular interest as this
parameter because of importance of the merging applied to real
finite photonic systems of finite size $LN$ where $L$ is the
period and $N$ is the number of elementary cells. The smallest
value of wave vector $\pi/LN$ responsible for quasi SP BIC defines
the asymptotical behavior of $Q_{\perp}$-factor over $N$ of the
quasi BIC mode because of leakage from surface of grating
\cite{Taghizadeh2017,Sadrieva2019}. We presented analytical theory
based on multipole decomposition of this radiation lose. For
merging of FW off-$\Gamma$ BIC and SP BIC at $\Gamma$-point for
limiting of the wave vector to zero a surface radiation is
completely suppressed leaving smaller radiation from the ends of
finite grating which decreases with the number of rod as $1/N^3$
\cite{Blaustein2007,Polishchuk,Asenjo2017,Bulgakov2019a}. That
transforms asymptotical behavior of $Q$-factor from standard law
$k_{x,z}^{-2}$  to super high $Q$-factor behavior $k_{x,z}^{-6}$
that justifies a terminology super BIC
\cite{Jin2019,Koshelev2019a,Kornovan2021,Kang2021}. Respectively
for finite grating we obtain the crossover of $Q$-factor from
standard square law $N^2$ towards cubic one $N^3$.

In addition we presented simple analytical theory based on generic
non Hermitian effective Hamiltonian (CMT model) accounting two
frequency bands of PhC which explains all above described
phenomena. That approach constitutes a important difference
compared to paper \cite{Kikkawa2020} in which CMT theory was
explored to consider ACR of two off-$\Gamma$ BICs. Owing to
generic form of Hamiltonian (\ref{H2}) the model is applicable to
arbitrary PhC system in which merging of BICs was observed
\cite{Jin2019,Hwang2021,Kang2021,Zhang2022}. The theory completely
agrees with multipole decomposition theory  and numerical Comsol
Multiphysics results.

\begin{acknowledgments}
We thankful Yi Xu and Zhanyuan Zhang for numerous fruitful
discussions. The work was supported by Russian Science Foundation
Grant No. 22-12-00070.
\end{acknowledgments}

\begin{thebibliography}{65}%
\makeatletter
\providecommand \@ifxundefined [1]{%
 \@ifx{#1\undefined}
}%
\providecommand \@ifnum [1]{%
 \ifnum #1\expandafter \@firstoftwo
 \else \expandafter \@secondoftwo
 \fi
}%
\providecommand \@ifx [1]{%
 \ifx #1\expandafter \@firstoftwo
 \else \expandafter \@secondoftwo
 \fi
}%
\providecommand \natexlab [1]{#1}%
\providecommand \enquote  [1]{``#1''}%
\providecommand \bibnamefont  [1]{#1}%
\providecommand \bibfnamefont [1]{#1}%
\providecommand \citenamefont [1]{#1}%
\providecommand \href@noop [0]{\@secondoftwo}%
\providecommand \href [0]{\begingroup \@sanitize@url \@href}%
\providecommand \@href[1]{\@@startlink{#1}\@@href}%
\providecommand \@@href[1]{\endgroup#1\@@endlink}%
\providecommand \@sanitize@url [0]{\catcode `\\12\catcode
`\$12\catcode
  `\&12\catcode `\#12\catcode `\^12\catcode `\_12\catcode `\%12\relax}%
\providecommand \@@startlink[1]{}%
\providecommand \@@endlink[0]{}%
\providecommand \url  [0]{\begingroup\@sanitize@url \@url }%
\providecommand \@url [1]{\endgroup\@href {#1}{\urlprefix }}%
\providecommand \urlprefix  [0]{URL }%
\providecommand \Eprint [0]{\href }%
\providecommand \doibase [0]{https://doi.org/}%
\providecommand \selectlanguage [0]{\@gobble}%
\providecommand \bibinfo  [0]{\@secondoftwo}%
\providecommand \bibfield  [0]{\@secondoftwo}%
\providecommand \translation [1]{[#1]}%
\providecommand \BibitemOpen [0]{}%
\providecommand \bibitemStop [0]{}%
\providecommand \bibitemNoStop [0]{.\EOS\space}%
\providecommand \EOS [0]{\spacefactor3000\relax}%
\providecommand \BibitemShut  [1]{\csname bibitem#1\endcsname}%
\let\auto@bib@innerbib\@empty
\bibitem [{\citenamefont {Vahala}(2003)}]{Vahala2003}%
  \BibitemOpen
  \bibfield  {author} {\bibinfo {author} {\bibfnamefont {K.~J.}\ \bibnamefont
  {Vahala}},\ }\bibfield  {title} {\bibinfo {title} {Optical microcavities},\
  }\href {https://doi.org/10.1038/nature01939} {\bibfield  {journal} {\bibinfo
  {journal} {Nature}\ }\textbf {\bibinfo {volume} {424}},\ \bibinfo {pages}
  {839} (\bibinfo {year} {2003})}\BibitemShut {NoStop}%
\bibitem [{\citenamefont {Braginsky}\ \emph {et~al.}(1989)\citenamefont
  {Braginsky}, \citenamefont {Gorodetsky},\ and\ \citenamefont
  {Ilchenko}}]{Braginsky1989}%
  \BibitemOpen
  \bibfield  {author} {\bibinfo {author} {\bibfnamefont {V.}~\bibnamefont
  {Braginsky}}, \bibinfo {author} {\bibfnamefont {M.}~\bibnamefont
  {Gorodetsky}},\ and\ \bibinfo {author} {\bibfnamefont {V.}~\bibnamefont
  {Ilchenko}},\ }\bibfield  {title} {\bibinfo {title} {Quality-factor and
  nonlinear properties of optical whispering-gallery modes},\ }\href
  {https://doi.org/10.1016/0375-9601(89)90912-2} {\bibfield  {journal}
  {\bibinfo  {journal} {Phys. Let. A}\ }\textbf {\bibinfo {volume} {137}},\
  \bibinfo {pages} {393} (\bibinfo {year} {1989})}\BibitemShut {NoStop}%
\bibitem [{\citenamefont {Gorodetsky}\ and\ \citenamefont
  {Ilchenko}(1999)}]{Gorodetsky1999}%
  \BibitemOpen
  \bibfield  {author} {\bibinfo {author} {\bibfnamefont {M.~L.}\ \bibnamefont
  {Gorodetsky}}\ and\ \bibinfo {author} {\bibfnamefont {V.~S.}\ \bibnamefont
  {Ilchenko}},\ }\bibfield  {title} {\bibinfo {title} {Optical microsphere
  resonators: optimal coupling to high-q whispering-gallery modes},\ }\href
  {https://doi.org/10.1364/josab.16.000147} {\bibfield  {journal} {\bibinfo
  {journal} {J. Opt. Soc. Am. B}\ }\textbf {\bibinfo {volume} {16}},\ \bibinfo
  {pages} {147} (\bibinfo {year} {1999})}\BibitemShut {NoStop}%
\bibitem [{\citenamefont {Ryckman}\ and\ \citenamefont
  {Weiss}(2012)}]{Ryckman2012}%
  \BibitemOpen
  \bibfield  {author} {\bibinfo {author} {\bibfnamefont {J.~D.}\ \bibnamefont
  {Ryckman}}\ and\ \bibinfo {author} {\bibfnamefont {S.~M.}\ \bibnamefont
  {Weiss}},\ }\bibfield  {title} {\bibinfo {title} {Low mode volume slotted
  photonic crystal single nanobeam cavity},\ }\href
  {https://doi.org/10.1063/1.4742749} {\bibfield  {journal} {\bibinfo
  {journal} {Applied Physics Letters}\ }\textbf {\bibinfo {volume} {101}},\
  \bibinfo {pages} {071104} (\bibinfo {year} {2012})}\BibitemShut {NoStop}%
\bibitem [{\citenamefont {Seidler}\ \emph {et~al.}(2013)\citenamefont
  {Seidler}, \citenamefont {Lister}, \citenamefont {Drechsler}, \citenamefont
  {Hofrichter},\ and\ \citenamefont {Stifferle}}]{Seidler2013}%
  \BibitemOpen
  \bibfield  {author} {\bibinfo {author} {\bibfnamefont {P.}~\bibnamefont
  {Seidler}}, \bibinfo {author} {\bibfnamefont {K.}~\bibnamefont {Lister}},
  \bibinfo {author} {\bibfnamefont {U.}~\bibnamefont {Drechsler}}, \bibinfo
  {author} {\bibfnamefont {J.}~\bibnamefont {Hofrichter}},\ and\ \bibinfo
  {author} {\bibfnamefont {T.}~\bibnamefont {Stifferle}},\ }\bibfield  {title}
  {\bibinfo {title} {Slotted photonic crystal nanobeam cavity with an ultrahigh
  quality factor-to-mode volume ratio},\ }\href
  {https://doi.org/10.1364/oe.21.032468} {\bibfield  {journal} {\bibinfo
  {journal} {Optics Express}\ }\textbf {\bibinfo {volume} {21}},\ \bibinfo
  {pages} {32468} (\bibinfo {year} {2013})}\BibitemShut {NoStop}%
\bibitem [{\citenamefont {Zhou}\ \emph {et~al.}(2019)\citenamefont {Zhou},
  \citenamefont {Zheng}, \citenamefont {Fang}, \citenamefont {Xu},\ and\
  \citenamefont {Majumdar}}]{Zhou2019}%
  \BibitemOpen
  \bibfield  {author} {\bibinfo {author} {\bibfnamefont {J.}~\bibnamefont
  {Zhou}}, \bibinfo {author} {\bibfnamefont {J.}~\bibnamefont {Zheng}},
  \bibinfo {author} {\bibfnamefont {Z.}~\bibnamefont {Fang}}, \bibinfo {author}
  {\bibfnamefont {P.}~\bibnamefont {Xu}},\ and\ \bibinfo {author}
  {\bibfnamefont {A.}~\bibnamefont {Majumdar}},\ }\bibfield  {title} {\bibinfo
  {title} {Ultra-low mode volume on-substrate silicon nanobeam cavity},\ }\href
  {https://doi.org/10.1364/oe.27.030692} {\bibfield  {journal} {\bibinfo
  {journal} {Optics Express}\ }\textbf {\bibinfo {volume} {27}},\ \bibinfo
  {pages} {30692} (\bibinfo {year} {2019})}\BibitemShut {NoStop}%
\bibitem [{\citenamefont {Hsu}\ \emph {et~al.}(2013{\natexlab{a}})\citenamefont
  {Hsu}, \citenamefont {Zhen}, \citenamefont {Chua}, \citenamefont {Johnson},
  \citenamefont {Joannopoulos},\ and\ \citenamefont
  {Solja{\v{c}}i{\'{c}}}}]{Hsu2016}%
  \BibitemOpen
  \bibfield  {author} {\bibinfo {author} {\bibfnamefont {C.~W.}\ \bibnamefont
  {Hsu}}, \bibinfo {author} {\bibfnamefont {B.}~\bibnamefont {Zhen}}, \bibinfo
  {author} {\bibfnamefont {S.-L.}\ \bibnamefont {Chua}}, \bibinfo {author}
  {\bibfnamefont {S.~G.}\ \bibnamefont {Johnson}}, \bibinfo {author}
  {\bibfnamefont {J.~D.}\ \bibnamefont {Joannopoulos}},\ and\ \bibinfo {author}
  {\bibfnamefont {M.}~\bibnamefont {Solja{\v{c}}i{\'{c}}}},\ }\bibfield
  {title} {\bibinfo {title} {Bloch surface eigenstates within the radiation
  continuum},\ }\href {https://doi.org/10.1038/lsa.2013.40} {\bibfield
  {journal} {\bibinfo  {journal} {Light: Science and Applications}\ }\textbf
  {\bibinfo {volume} {2}},\ \bibinfo {pages} {e84} (\bibinfo {year}
  {2013}{\natexlab{a}})}\BibitemShut {NoStop}%
\bibitem [{\citenamefont {Azzam}\ and\ \citenamefont
  {Kildishev}(2020)}]{Azzam2020}%
  \BibitemOpen
  \bibfield  {author} {\bibinfo {author} {\bibfnamefont {S.~I.}\ \bibnamefont
  {Azzam}}\ and\ \bibinfo {author} {\bibfnamefont {A.~V.}\ \bibnamefont
  {Kildishev}},\ }\bibfield  {title} {\bibinfo {title} {Photonic bound states
  in the continuum: From basics to applications},\ }\href
  {https://doi.org/10.1002/adom.202001469} {\bibfield  {journal} {\bibinfo
  {journal} {Adv. Opt. Mater.}\ }\textbf {\bibinfo {volume} {9}},\ \bibinfo
  {pages} {2001469} (\bibinfo {year} {2020})}\BibitemShut {NoStop}%
\bibitem [{\citenamefont {Huang}\ \emph {et~al.}(2020)\citenamefont {Huang},
  \citenamefont {Xu}, \citenamefont {Woolley},\ and\ \citenamefont
  {Miroshnichenko}}]{Huang2020}%
  \BibitemOpen
  \bibfield  {author} {\bibinfo {author} {\bibfnamefont {L.}~\bibnamefont
  {Huang}}, \bibinfo {author} {\bibfnamefont {L.}~\bibnamefont {Xu}}, \bibinfo
  {author} {\bibfnamefont {M.}~\bibnamefont {Woolley}},\ and\ \bibinfo {author}
  {\bibfnamefont {A.}~\bibnamefont {Miroshnichenko}},\ }\bibfield  {title}
  {\bibinfo {title} {Trends in quantum nanophotonics},\ }\href
  {https://doi.org/10.1002/qute.201900126} {\bibfield  {journal} {\bibinfo
  {journal} {Adv. Quantum Technologies}\ }\textbf {\bibinfo {volume} {3}},\
  \bibinfo {pages} {1900126} (\bibinfo {year} {2020})}\BibitemShut {NoStop}%
\bibitem [{\citenamefont {Hu}\ \emph {et~al.}(2020)\citenamefont {Hu},
  \citenamefont {Yuan},\ and\ \citenamefont {Lu}}]{Hu2020}%
  \BibitemOpen
  \bibfield  {author} {\bibinfo {author} {\bibfnamefont {Z.}~\bibnamefont
  {Hu}}, \bibinfo {author} {\bibfnamefont {L.}~\bibnamefont {Yuan}},\ and\
  \bibinfo {author} {\bibfnamefont {Y.~Y.}\ \bibnamefont {Lu}},\ }\bibfield
  {title} {\bibinfo {title} {Resonant field enhancement near bound states in
  the continuum on periodic structures},\ }\href
  {https://doi.org/10.1103/physreva.101.043825} {\bibfield  {journal} {\bibinfo
   {journal} {Phys. Rev. A}\ }\textbf {\bibinfo {volume} {101}},\ \bibinfo
  {pages} {043825} (\bibinfo {year} {2020})}\BibitemShut {NoStop}%
\bibitem [{\citenamefont {Joseph}\ \emph {et~al.}(2021)\citenamefont {Joseph},
  \citenamefont {Pandey}, \citenamefont {Sarkar},\ and\ \citenamefont
  {Joseph}}]{Joseph2021}%
  \BibitemOpen
  \bibfield  {author} {\bibinfo {author} {\bibfnamefont {S.}~\bibnamefont
  {Joseph}}, \bibinfo {author} {\bibfnamefont {S.}~\bibnamefont {Pandey}},
  \bibinfo {author} {\bibfnamefont {S.}~\bibnamefont {Sarkar}},\ and\ \bibinfo
  {author} {\bibfnamefont {J.}~\bibnamefont {Joseph}},\ }\bibfield  {title}
  {\bibinfo {title} {Bound states in the continuum in resonant nanostructures:
  an overview of engineered materials for tailored applications},\ }\href
  {https://doi.org/10.1515/nanoph-2021-0387} {\bibfield  {journal} {\bibinfo
  {journal} {Nanophotonics}\ }\textbf {\bibinfo {volume} {10}},\ \bibinfo
  {pages} {4175} (\bibinfo {year} {2021})}\BibitemShut {NoStop}%
\bibitem [{\citenamefont {Koshelev}\ \emph {et~al.}(2021)\citenamefont
  {Koshelev}, \citenamefont {Sadrieva}, \citenamefont {Shcherbakov},
  \citenamefont {Kivshar},\ and\ \citenamefont {Bogdanov}}]{Koshelev2021}%
  \BibitemOpen
  \bibfield  {author} {\bibinfo {author} {\bibfnamefont {K.}~\bibnamefont
  {Koshelev}}, \bibinfo {author} {\bibfnamefont {Z.}~\bibnamefont {Sadrieva}},
  \bibinfo {author} {\bibfnamefont {A.}~\bibnamefont {Shcherbakov}}, \bibinfo
  {author} {\bibfnamefont {Y.}~\bibnamefont {Kivshar}},\ and\ \bibinfo {author}
  {\bibfnamefont {A.}~\bibnamefont {Bogdanov}},\ }\bibfield  {title} {\bibinfo
  {title} {Bound states of the continuum in photonic structures},\ }\bibfield
  {journal} {\bibinfo  {journal} {Physics-Uspekhi}\ }\textbf {\bibinfo {volume}
  {65}},\ \href {https://doi.org/10.3367/ufne.2021.12.039120}
  {10.3367/ufne.2021.12.039120} (\bibinfo {year} {2021})\BibitemShut {NoStop}%
\bibitem [{\citenamefont {Hu}\ \emph {et~al.}(2023)\citenamefont {Hu},
  \citenamefont {Xie}, \citenamefont {Song}, \citenamefont {Chen},
  \citenamefont {Xiang}, \citenamefont {Han},\ and\ \citenamefont
  {Zi}}]{Hu2023}%
  \BibitemOpen
  \bibfield  {author} {\bibinfo {author} {\bibfnamefont {P.}~\bibnamefont
  {Hu}}, \bibinfo {author} {\bibfnamefont {C.}~\bibnamefont {Xie}}, \bibinfo
  {author} {\bibfnamefont {Q.}~\bibnamefont {Song}}, \bibinfo {author}
  {\bibfnamefont {A.}~\bibnamefont {Chen}}, \bibinfo {author} {\bibfnamefont
  {H.}~\bibnamefont {Xiang}}, \bibinfo {author} {\bibfnamefont
  {D.}~\bibnamefont {Han}},\ and\ \bibinfo {author} {\bibfnamefont
  {J.}~\bibnamefont {Zi}},\ }\bibfield  {title} {\bibinfo {title} {Bound states
  in the continuum based on the total internal reflection of bloch waves},\
  }\bibfield  {journal} {\bibinfo  {journal} {National Science Review}\
  }\textbf {\bibinfo {volume} {10}},\ \href
  {https://doi.org/10.1093/nsr/nwac043} {10.1093/nsr/nwac043} (\bibinfo {year}
  {2023})\BibitemShut {NoStop}%
\bibitem [{\citenamefont {Hsu}\ \emph {et~al.}(2013{\natexlab{b}})\citenamefont
  {Hsu}, \citenamefont {Zhen}, \citenamefont {Lee}, \citenamefont {Johnson},
  \citenamefont {Joannopoulos},\ and\ \citenamefont
  {Solja{\v{c}}i{\'c}}}]{Hsu2013}%
  \BibitemOpen
  \bibfield  {author} {\bibinfo {author} {\bibfnamefont {C.~W.}\ \bibnamefont
  {Hsu}}, \bibinfo {author} {\bibfnamefont {B.}~\bibnamefont {Zhen}}, \bibinfo
  {author} {\bibfnamefont {J.}~\bibnamefont {Lee}}, \bibinfo {author}
  {\bibfnamefont {S.~G.}\ \bibnamefont {Johnson}}, \bibinfo {author}
  {\bibfnamefont {J.~D.}\ \bibnamefont {Joannopoulos}},\ and\ \bibinfo {author}
  {\bibfnamefont {M.}~\bibnamefont {Solja{\v{c}}i{\'c}}},\ }\bibfield  {title}
  {\bibinfo {title} {Observation of trapped light within the radiation
  continuum},\ }\href {https://doi.org/10.1038/nature12289} {\bibfield
  {journal} {\bibinfo  {journal} {Nature}\ }\textbf {\bibinfo {volume} {499}},\
  \bibinfo {pages} {188} (\bibinfo {year} {2013}{\natexlab{b}})}\BibitemShut
  {NoStop}%
\bibitem [{\citenamefont {Bulgakov}\ and\ \citenamefont
  {Sadreev}(2017{\natexlab{a}})}]{Bulgakov2017}%
  \BibitemOpen
  \bibfield  {author} {\bibinfo {author} {\bibfnamefont {E.}~\bibnamefont
  {Bulgakov}}\ and\ \bibinfo {author} {\bibfnamefont {A.}~\bibnamefont
  {Sadreev}},\ }\bibfield  {title} {\bibinfo {title} {Bound states in the
  continuum with high orbital angular momentum in a dielectric rod with
  periodically modulated permittivity},\ }\href
  {https://doi.org/10.1103/physreva.96.013841} {\bibfield  {journal} {\bibinfo
  {journal} {Phys. Rev. A}\ }\textbf {\bibinfo {volume} {96}},\ \bibinfo
  {pages} {013841} (\bibinfo {year} {2017}{\natexlab{a}})}\BibitemShut
  {NoStop}%
\bibitem [{\citenamefont {Koshelev}\ \emph {et~al.}(2019)\citenamefont
  {Koshelev}, \citenamefont {Favraud}, \citenamefont {Bogdanov}, \citenamefont
  {Kivshar},\ and\ \citenamefont {Fratalocchi}}]{Koshelev2019}%
  \BibitemOpen
  \bibfield  {author} {\bibinfo {author} {\bibfnamefont {K.}~\bibnamefont
  {Koshelev}}, \bibinfo {author} {\bibfnamefont {G.}~\bibnamefont {Favraud}},
  \bibinfo {author} {\bibfnamefont {A.}~\bibnamefont {Bogdanov}}, \bibinfo
  {author} {\bibfnamefont {Y.}~\bibnamefont {Kivshar}},\ and\ \bibinfo {author}
  {\bibfnamefont {A.}~\bibnamefont {Fratalocchi}},\ }\bibfield  {title}
  {\bibinfo {title} {Nonradiating photonics with resonant dielectric
  nanostructures},\ }\href {https://doi.org/10.1515/nanoph-2019-0024}
  {\bibfield  {journal} {\bibinfo  {journal} {Nanophotonics}\ }\textbf
  {\bibinfo {volume} {8}},\ \bibinfo {pages} {725} (\bibinfo {year}
  {2019})}\BibitemShut {NoStop}%
\bibitem [{\citenamefont {Taghizadeh}\ and\ \citenamefont
  {Chung}(2017)}]{Taghizadeh2017}%
  \BibitemOpen
  \bibfield  {author} {\bibinfo {author} {\bibfnamefont {A.}~\bibnamefont
  {Taghizadeh}}\ and\ \bibinfo {author} {\bibfnamefont {I.-S.}\ \bibnamefont
  {Chung}},\ }\bibfield  {title} {\bibinfo {title} {Quasi bound states in the
  continuum with few unit cells of photonic crystal slab},\ }\href
  {https://doi.org/10.1063/1.4990753} {\bibfield  {journal} {\bibinfo
  {journal} {Appl. Phys. Lett.}\ }\textbf {\bibinfo {volume} {111}},\ \bibinfo
  {pages} {031114} (\bibinfo {year} {2017})}\BibitemShut {NoStop}%
\bibitem [{\citenamefont {Bulgakov}\ and\ \citenamefont
  {Sadreev}(2017{\natexlab{b}})}]{Bulgakov2017b}%
  \BibitemOpen
  \bibfield  {author} {\bibinfo {author} {\bibfnamefont {E.}~\bibnamefont
  {Bulgakov}}\ and\ \bibinfo {author} {\bibfnamefont {A.}~\bibnamefont
  {Sadreev}},\ }\bibfield  {title} {\bibinfo {title} {Propagating bloch bound
  states with orbital angular momentum above the light line in the array of
  dielectric spheres},\ }\href {https://doi.org/10.1364/josaa.34.000949}
  {\bibfield  {journal} {\bibinfo  {journal} {J. Opt. Soc. Am. A}\ }\textbf
  {\bibinfo {volume} {34}},\ \bibinfo {pages} {949} (\bibinfo {year}
  {2017}{\natexlab{b}})}\BibitemShut {NoStop}%
\bibitem [{\citenamefont {Sadrieva}\ \emph {et~al.}(2019)\citenamefont
  {Sadrieva}, \citenamefont {Belyakov}, \citenamefont {Balezin}, \citenamefont
  {Kapitanova}, \citenamefont {Nenasheva}, \citenamefont {Sadreev},\ and\
  \citenamefont {Bogdanov}}]{Sadrieva2019}%
  \BibitemOpen
  \bibfield  {author} {\bibinfo {author} {\bibfnamefont {Z.~F.}\ \bibnamefont
  {Sadrieva}}, \bibinfo {author} {\bibfnamefont {M.~A.}\ \bibnamefont
  {Belyakov}}, \bibinfo {author} {\bibfnamefont {M.~A.}\ \bibnamefont
  {Balezin}}, \bibinfo {author} {\bibfnamefont {P.~V.}\ \bibnamefont
  {Kapitanova}}, \bibinfo {author} {\bibfnamefont {E.~A.}\ \bibnamefont
  {Nenasheva}}, \bibinfo {author} {\bibfnamefont {A.~F.}\ \bibnamefont
  {Sadreev}},\ and\ \bibinfo {author} {\bibfnamefont {A.~A.}\ \bibnamefont
  {Bogdanov}},\ }\bibfield  {title} {\bibinfo {title} {Experimental observation
  of a symmetry-protected bound state in the continuum in a chain of dielectric
  disks},\ }\href {https://doi.org/10.1103/physreva.99.053804} {\bibfield
  {journal} {\bibinfo  {journal} {Phys. Rev. A}\ }\textbf {\bibinfo {volume}
  {99}},\ \bibinfo {pages} {053804} (\bibinfo {year} {2019})}\BibitemShut
  {NoStop}%
\bibitem [{\citenamefont {Polishchuk}\ \emph {et~al.}(2017)\citenamefont
  {Polishchuk}, \citenamefont {Anastasiev}, \citenamefont {Tsyvkunova},
  \citenamefont {Gozman}, \citenamefont {Solov'ov},\ and\ \citenamefont
  {Polishchuk}}]{Polishchuk}%
  \BibitemOpen
  \bibfield  {author} {\bibinfo {author} {\bibfnamefont {I.~Y.}\ \bibnamefont
  {Polishchuk}}, \bibinfo {author} {\bibfnamefont {A.~A.}\ \bibnamefont
  {Anastasiev}}, \bibinfo {author} {\bibfnamefont {E.~A.}\ \bibnamefont
  {Tsyvkunova}}, \bibinfo {author} {\bibfnamefont {M.~I.}\ \bibnamefont
  {Gozman}}, \bibinfo {author} {\bibfnamefont {S.~V.}\ \bibnamefont
  {Solov'ov}},\ and\ \bibinfo {author} {\bibfnamefont {Y.~I.}\ \bibnamefont
  {Polishchuk}},\ }\bibfield  {title} {\bibinfo {title} {Guided modes in the
  plane array of optical waveguides},\ }\href
  {https://doi.org/10.1103/physreva.95.053847} {\bibfield  {journal} {\bibinfo
  {journal} {Physical Review A}\ }\textbf {\bibinfo {volume} {95}},\ \bibinfo
  {pages} {053847} (\bibinfo {year} {2017})}\BibitemShut {NoStop}%
\bibitem [{\citenamefont {Sidorenko}\ \emph {et~al.}(2021)\citenamefont
  {Sidorenko}, \citenamefont {Sergaeva}, \citenamefont {Sadrieva},
  \citenamefont {Roques-Carmes}, \citenamefont {Muraev}, \citenamefont
  {Maksimov},\ and\ \citenamefont {Bogdanov}}]{Sidorenko2021}%
  \BibitemOpen
  \bibfield  {author} {\bibinfo {author} {\bibfnamefont {M.}~\bibnamefont
  {Sidorenko}}, \bibinfo {author} {\bibfnamefont {O.}~\bibnamefont {Sergaeva}},
  \bibinfo {author} {\bibfnamefont {Z.}~\bibnamefont {Sadrieva}}, \bibinfo
  {author} {\bibfnamefont {C.}~\bibnamefont {Roques-Carmes}}, \bibinfo {author}
  {\bibfnamefont {P.}~\bibnamefont {Muraev}}, \bibinfo {author} {\bibfnamefont
  {D.}~\bibnamefont {Maksimov}},\ and\ \bibinfo {author} {\bibfnamefont
  {A.}~\bibnamefont {Bogdanov}},\ }\bibfield  {title} {\bibinfo {title}
  {Observation of an accidental bound state in the continuum in a chain of
  dielectric disks},\ }\href {https://doi.org/10.1103/physrevapplied.15.034041}
  {\bibfield  {journal} {\bibinfo  {journal} {Phys. Rev. Appl.}\ }\textbf
  {\bibinfo {volume} {15}},\ \bibinfo {pages} {034041} (\bibinfo {year}
  {2021})}\BibitemShut {NoStop}%
\bibitem [{\citenamefont {Zhang}\ \emph {et~al.}()\citenamefont {Zhang},
  \citenamefont {Bulgakov}, \citenamefont {Pichugin}, \citenamefont {Sadreev},
  \citenamefont {Xu},\ and\ \citenamefont {Qin}}]{Zhang2022}%
  \BibitemOpen
  \bibfield  {author} {\bibinfo {author} {\bibfnamefont {Z.}~\bibnamefont
  {Zhang}}, \bibinfo {author} {\bibfnamefont {E.}~\bibnamefont {Bulgakov}},
  \bibinfo {author} {\bibfnamefont {K.}~\bibnamefont {Pichugin}}, \bibinfo
  {author} {\bibfnamefont {A.}~\bibnamefont {Sadreev}}, \bibinfo {author}
  {\bibfnamefont {Y.}~\bibnamefont {Xu}},\ and\ \bibinfo {author}
  {\bibfnamefont {Y.}~\bibnamefont {Qin}},\ }\bibfield  {title} {\bibinfo
  {title} {Super quasi-bound state in the continuum},\ }\href@noop {} {\
  }\Eprint {https://arxiv.org/abs/2211.03675v2} {2211.03675v2} \BibitemShut
  {NoStop}%
\bibitem [{\citenamefont {Ni}\ \emph {et~al.}(2017)\citenamefont {Ni},
  \citenamefont {Jin}, \citenamefont {Peng},\ and\ \citenamefont
  {Li}}]{Ni2017}%
  \BibitemOpen
  \bibfield  {author} {\bibinfo {author} {\bibfnamefont {L.}~\bibnamefont
  {Ni}}, \bibinfo {author} {\bibfnamefont {J.}~\bibnamefont {Jin}}, \bibinfo
  {author} {\bibfnamefont {C.}~\bibnamefont {Peng}},\ and\ \bibinfo {author}
  {\bibfnamefont {Z.}~\bibnamefont {Li}},\ }\bibfield  {title} {\bibinfo
  {title} {Analytical and statistical investigation on structural fluctuations
  induced radiation in photonic crystal slabs},\ }\href
  {https://doi.org/10.1364/oe.25.005580} {\bibfield  {journal} {\bibinfo
  {journal} {Optics Express}\ }\textbf {\bibinfo {volume} {25}},\ \bibinfo
  {pages} {5580} (\bibinfo {year} {2017})}\BibitemShut {NoStop}%
\bibitem [{\citenamefont {Maslova}\ \emph {et~al.}(2021)\citenamefont
  {Maslova}, \citenamefont {Rybin}, \citenamefont {Bogdanov},\ and\
  \citenamefont {Sadrieva}}]{Maslova2021}%
  \BibitemOpen
  \bibfield  {author} {\bibinfo {author} {\bibfnamefont {E.~E.}\ \bibnamefont
  {Maslova}}, \bibinfo {author} {\bibfnamefont {M.~V.}\ \bibnamefont {Rybin}},
  \bibinfo {author} {\bibfnamefont {A.~A.}\ \bibnamefont {Bogdanov}},\ and\
  \bibinfo {author} {\bibfnamefont {Z.~F.}\ \bibnamefont {Sadrieva}},\
  }\bibfield  {title} {\bibinfo {title} {Bound states in the continuum in
  periodic structures with structural disorder},\ }\href
  {https://doi.org/10.1515/nanoph-2021-0475} {\bibfield  {journal} {\bibinfo
  {journal} {Nanophotonics}\ }\textbf {\bibinfo {volume} {10}},\ \bibinfo
  {pages} {4313} (\bibinfo {year} {2021})}\BibitemShut {NoStop}%
\bibitem [{\citenamefont {Jin}\ \emph {et~al.}(2019)\citenamefont {Jin},
  \citenamefont {Yin}, \citenamefont {Ni}, \citenamefont
  {Solja{\v{c}}i{\'{c}}}, \citenamefont {Zhen},\ and\ \citenamefont
  {Peng}}]{Jin2019}%
  \BibitemOpen
  \bibfield  {author} {\bibinfo {author} {\bibfnamefont {J.}~\bibnamefont
  {Jin}}, \bibinfo {author} {\bibfnamefont {X.}~\bibnamefont {Yin}}, \bibinfo
  {author} {\bibfnamefont {L.}~\bibnamefont {Ni}}, \bibinfo {author}
  {\bibfnamefont {M.}~\bibnamefont {Solja{\v{c}}i{\'{c}}}}, \bibinfo {author}
  {\bibfnamefont {B.}~\bibnamefont {Zhen}},\ and\ \bibinfo {author}
  {\bibfnamefont {C.}~\bibnamefont {Peng}},\ }\bibfield  {title} {\bibinfo
  {title} {Topologically enabled ultrahigh-q guided resonances robust to
  out-of-plane scattering},\ }\href {https://doi.org/10.1038/s41586-019-1664-7}
  {\bibfield  {journal} {\bibinfo  {journal} {Nature}\ }\textbf {\bibinfo
  {volume} {574}},\ \bibinfo {pages} {501} (\bibinfo {year}
  {2019})}\BibitemShut {NoStop}%
\bibitem [{\citenamefont {Hwang}\ \emph {et~al.}(2021)\citenamefont {Hwang},
  \citenamefont {Lee}, \citenamefont {Kim}, \citenamefont {Jeong},
  \citenamefont {Kwon}, \citenamefont {Koshelev}, \citenamefont {Kivshar},\
  and\ \citenamefont {Park}}]{Hwang2021}%
  \BibitemOpen
  \bibfield  {author} {\bibinfo {author} {\bibfnamefont {M.-S.}\ \bibnamefont
  {Hwang}}, \bibinfo {author} {\bibfnamefont {H.-C.}\ \bibnamefont {Lee}},
  \bibinfo {author} {\bibfnamefont {K.-H.}\ \bibnamefont {Kim}}, \bibinfo
  {author} {\bibfnamefont {K.-Y.}\ \bibnamefont {Jeong}}, \bibinfo {author}
  {\bibfnamefont {S.-H.}\ \bibnamefont {Kwon}}, \bibinfo {author}
  {\bibfnamefont {K.}~\bibnamefont {Koshelev}}, \bibinfo {author}
  {\bibfnamefont {Y.}~\bibnamefont {Kivshar}},\ and\ \bibinfo {author}
  {\bibfnamefont {H.-G.}\ \bibnamefont {Park}},\ }\bibfield  {title} {\bibinfo
  {title} {Ultralow-threshold laser using super-bound states in the
  continuum},\ }\bibfield  {journal} {\bibinfo  {journal} {Nature
  Communications}\ }\textbf {\bibinfo {volume} {12}},\ \href
  {https://doi.org/10.1038/s41467-021-24502-0} {10.1038/s41467-021-24502-0}
  (\bibinfo {year} {2021})\BibitemShut {NoStop}%
\bibitem [{\citenamefont {Kang}\ \emph {et~al.}(2021)\citenamefont {Kang},
  \citenamefont {Zhang}, \citenamefont {Xiao},\ and\ \citenamefont
  {Xu}}]{Kang2021}%
  \BibitemOpen
  \bibfield  {author} {\bibinfo {author} {\bibfnamefont {M.}~\bibnamefont
  {Kang}}, \bibinfo {author} {\bibfnamefont {S.}~\bibnamefont {Zhang}},
  \bibinfo {author} {\bibfnamefont {M.}~\bibnamefont {Xiao}},\ and\ \bibinfo
  {author} {\bibfnamefont {H.}~\bibnamefont {Xu}},\ }\bibfield  {title}
  {\bibinfo {title} {Merging bound states in the continuum at off-high symmetry
  points},\ }\href {https://doi.org/10.1103/physrevlett.126.117402} {\bibfield
  {journal} {\bibinfo  {journal} {Phys. Rev. Lett.}\ }\textbf {\bibinfo
  {volume} {126}},\ \bibinfo {pages} {117402} (\bibinfo {year}
  {2021})}\BibitemShut {NoStop}%
\bibitem [{\citenamefont {Bulgakov}\ \emph {et~al.}(2022)\citenamefont
  {Bulgakov}, \citenamefont {Pilipchuk},\ and\ \citenamefont
  {Sadreev}}]{Bulgakov2022}%
  \BibitemOpen
  \bibfield  {author} {\bibinfo {author} {\bibfnamefont {E.}~\bibnamefont
  {Bulgakov}}, \bibinfo {author} {\bibfnamefont {A.}~\bibnamefont
  {Pilipchuk}},\ and\ \bibinfo {author} {\bibfnamefont {A.}~\bibnamefont
  {Sadreev}},\ }\bibfield  {title} {\bibinfo {title} {Desktop laboratory of
  bound states in the continuum in metallic waveguide with dielectric
  cavities},\ }\href {https://doi.org/10.1103/physrevb.106.075304} {\bibfield
  {journal} {\bibinfo  {journal} {Phys. Rev. B}\ }\textbf {\bibinfo {volume}
  {106}},\ \bibinfo {pages} {075304} (\bibinfo {year} {2022})}\BibitemShut
  {NoStop}%
\bibitem [{\citenamefont {Huang}\ \emph {et~al.}(2022)\citenamefont {Huang},
  \citenamefont {Jia}, \citenamefont {Chiang}, \citenamefont {Huang},
  \citenamefont {Shen}, \citenamefont {Deng}, \citenamefont {Yang},
  \citenamefont {Powell}, \citenamefont {Li},\ and\ \citenamefont
  {Miroshnichenko}}]{Huang2022}%
  \BibitemOpen
  \bibfield  {author} {\bibinfo {author} {\bibfnamefont {L.}~\bibnamefont
  {Huang}}, \bibinfo {author} {\bibfnamefont {B.}~\bibnamefont {Jia}}, \bibinfo
  {author} {\bibfnamefont {Y.~K.}\ \bibnamefont {Chiang}}, \bibinfo {author}
  {\bibfnamefont {S.}~\bibnamefont {Huang}}, \bibinfo {author} {\bibfnamefont
  {C.}~\bibnamefont {Shen}}, \bibinfo {author} {\bibfnamefont {F.}~\bibnamefont
  {Deng}}, \bibinfo {author} {\bibfnamefont {T.}~\bibnamefont {Yang}}, \bibinfo
  {author} {\bibfnamefont {D.~A.}\ \bibnamefont {Powell}}, \bibinfo {author}
  {\bibfnamefont {Y.}~\bibnamefont {Li}},\ and\ \bibinfo {author}
  {\bibfnamefont {A.~E.}\ \bibnamefont {Miroshnichenko}},\ }\bibfield  {title}
  {\bibinfo {title} {Topological supercavity resonances in the finite system},\
  }\href {https://doi.org/10.1002/advs.202200257} {\bibfield  {journal}
  {\bibinfo  {journal} {Advanced Science}\ }\textbf {\bibinfo {volume} {9}},\
  \bibinfo {pages} {2200257} (\bibinfo {year} {2022})}\BibitemShut {NoStop}%
\bibitem [{\citenamefont {Huang}\ \emph {et~al.}(2023)\citenamefont {Huang},
  \citenamefont {Xu}, \citenamefont {Powell}, \citenamefont {Padilla},\ and\
  \citenamefont {Miroshnichenko}}]{Huang2023}%
  \BibitemOpen
  \bibfield  {author} {\bibinfo {author} {\bibfnamefont {L.}~\bibnamefont
  {Huang}}, \bibinfo {author} {\bibfnamefont {L.}~\bibnamefont {Xu}}, \bibinfo
  {author} {\bibfnamefont {D.}~\bibnamefont {Powell}}, \bibinfo {author}
  {\bibfnamefont {W.}~\bibnamefont {Padilla}},\ and\ \bibinfo {author}
  {\bibfnamefont {A.}~\bibnamefont {Miroshnichenko}},\ }\bibfield  {title}
  {\bibinfo {title} {Resonant leaky modes in all-dielectric metasystems:
  Fundamentals and applications},\ }\href
  {https://doi.org/10.1016/j.physrep.2023.01.001} {\bibfield  {journal}
  {\bibinfo  {journal} {Phys. Rep.}\ }\textbf {\bibinfo {volume} {1008}},\
  \bibinfo {pages} {1} (\bibinfo {year} {2023})}\BibitemShut {NoStop}%
\bibitem [{\citenamefont {Wang}\ \emph {et~al.}(2023)\citenamefont {Wang},
  \citenamefont {Liu}, \citenamefont {Li}, \citenamefont {Liu}, \citenamefont
  {Zhang},\ and\ \citenamefont {Zhang}}]{Wang2023}%
  \BibitemOpen
  \bibfield  {author} {\bibinfo {author} {\bibfnamefont {K.}~\bibnamefont
  {Wang}}, \bibinfo {author} {\bibfnamefont {H.}~\bibnamefont {Liu}}, \bibinfo
  {author} {\bibfnamefont {Z.}~\bibnamefont {Li}}, \bibinfo {author}
  {\bibfnamefont {M.}~\bibnamefont {Liu}}, \bibinfo {author} {\bibfnamefont
  {Y.}~\bibnamefont {Zhang}},\ and\ \bibinfo {author} {\bibfnamefont
  {H.}~\bibnamefont {Zhang}},\ }\bibfield  {title} {\bibinfo {title}
  {All-dielectric metasurface-based multimode sensing with symmetry-protected
  and accidental bound states in the continuum},\ }\href
  {https://doi.org/10.1016/j.rinp.2023.106276} {\bibfield  {journal} {\bibinfo
  {journal} {Results in Physics}\ }\textbf {\bibinfo {volume} {46}},\ \bibinfo
  {pages} {106276} (\bibinfo {year} {2023})}\BibitemShut {NoStop}%
\bibitem [{\citenamefont {Shubin}(2023)}]{Shubin2023}%
  \BibitemOpen
  \bibfield  {author} {\bibinfo {author} {\bibfnamefont {N.~M.}\ \bibnamefont
  {Shubin}},\ }\bibfield  {title} {\bibinfo {title} {Algebraic approach to
  annihilation and repulsion of bound states in the continuum in finite
  systems},\ }\href {https://doi.org/10.1063/5.0142892} {\bibfield  {journal}
  {\bibinfo  {journal} {J. Math. Phys.}\ }\textbf {\bibinfo {volume} {64}},\
  \bibinfo {pages} {042103} (\bibinfo {year} {2023})}\BibitemShut {NoStop}%
\bibitem [{\citenamefont {Barkaoui}\ \emph {et~al.}(2023)\citenamefont
  {Barkaoui}, \citenamefont {Du}, \citenamefont {Chen}, \citenamefont {Xiao},\
  and\ \citenamefont {Song}}]{Barkaoui2023}%
  \BibitemOpen
  \bibfield  {author} {\bibinfo {author} {\bibfnamefont {H.}~\bibnamefont
  {Barkaoui}}, \bibinfo {author} {\bibfnamefont {K.}~\bibnamefont {Du}},
  \bibinfo {author} {\bibfnamefont {Y.}~\bibnamefont {Chen}}, \bibinfo {author}
  {\bibfnamefont {S.}~\bibnamefont {Xiao}},\ and\ \bibinfo {author}
  {\bibfnamefont {Q.}~\bibnamefont {Song}},\ }\bibfield  {title} {\bibinfo
  {title} {Merged bound states in the continuum for giant superchiral field and
  chiral mode splitting},\ }\href {https://doi.org/10.1103/physrevb.107.045305}
  {\bibfield  {journal} {\bibinfo  {journal} {Phys. Rev. B}\ }\textbf {\bibinfo
  {volume} {107}},\ \bibinfo {pages} {045305} (\bibinfo {year}
  {2023})}\BibitemShut {NoStop}%
\bibitem [{\citenamefont {Fan}\ \emph {et~al.}()\citenamefont {Fan},
  \citenamefont {Xue}, \citenamefont {Xing}, \citenamefont {Lu}, \citenamefont
  {Xu}, \citenamefont {Gu}, \citenamefont {Han},\ and\ \citenamefont
  {Cong}}]{Fan2023}%
  \BibitemOpen
  \bibfield  {author} {\bibinfo {author} {\bibfnamefont {J.}~\bibnamefont
  {Fan}}, \bibinfo {author} {\bibfnamefont {Z.}~\bibnamefont {Xue}}, \bibinfo
  {author} {\bibfnamefont {H.}~\bibnamefont {Xing}}, \bibinfo {author}
  {\bibfnamefont {D.}~\bibnamefont {Lu}}, \bibinfo {author} {\bibfnamefont
  {G.}~\bibnamefont {Xu}}, \bibinfo {author} {\bibfnamefont {J.}~\bibnamefont
  {Gu}}, \bibinfo {author} {\bibfnamefont {J.}~\bibnamefont {Han}},\ and\
  \bibinfo {author} {\bibfnamefont {L.}~\bibnamefont {Cong}},\ }\bibfield
  {title} {\bibinfo {title} {Hybrid bound states in the continuum in terahertz
  metasurfaces},\ }\href@noop {} {\ }\Eprint
  {https://arxiv.org/abs/2303.12264v1} {2303.12264v1} \BibitemShut {NoStop}%
\bibitem [{\citenamefont {Zhang}\ \emph {et~al.}(2023)\citenamefont {Zhang},
  \citenamefont {Zhang}, \citenamefont {Chen}, \citenamefont {Duan},
  \citenamefont {Li}, \citenamefont {Shi}, \citenamefont {Zi},\ and\
  \citenamefont {Zhang}}]{Zhang2023}%
  \BibitemOpen
  \bibfield  {author} {\bibinfo {author} {\bibfnamefont {H.}~\bibnamefont
  {Zhang}}, \bibinfo {author} {\bibfnamefont {W.}~\bibnamefont {Zhang}},
  \bibinfo {author} {\bibfnamefont {S.}~\bibnamefont {Chen}}, \bibinfo {author}
  {\bibfnamefont {P.}~\bibnamefont {Duan}}, \bibinfo {author} {\bibfnamefont
  {J.}~\bibnamefont {Li}}, \bibinfo {author} {\bibfnamefont {L.}~\bibnamefont
  {Shi}}, \bibinfo {author} {\bibfnamefont {J.}~\bibnamefont {Zi}},\ and\
  \bibinfo {author} {\bibfnamefont {X.}~\bibnamefont {Zhang}},\ }\bibfield
  {title} {\bibinfo {title} {Experimental observation of vector bound states in
  the continuum},\ }\href {https://doi.org/10.1002/adom.202203118} {\bibfield
  {journal} {\bibinfo  {journal} {Advanced Optical Materials}\ ,\ \bibinfo
  {pages} {2203118}} (\bibinfo {year} {2023})}\BibitemShut {NoStop}%
\bibitem [{\citenamefont {Qin}\ \emph {et~al.}(2023)\citenamefont {Qin},
  \citenamefont {Su}, \citenamefont {Liu}, \citenamefont {Zeng}, \citenamefont
  {Tang}, \citenamefont {Li}, \citenamefont {Shi}, \citenamefont {Huang},
  \citenamefont {Qiu},\ and\ \citenamefont {Song}}]{Qin2023}%
  \BibitemOpen
  \bibfield  {author} {\bibinfo {author} {\bibfnamefont {H.}~\bibnamefont
  {Qin}}, \bibinfo {author} {\bibfnamefont {Z.}~\bibnamefont {Su}}, \bibinfo
  {author} {\bibfnamefont {M.}~\bibnamefont {Liu}}, \bibinfo {author}
  {\bibfnamefont {Y.}~\bibnamefont {Zeng}}, \bibinfo {author} {\bibfnamefont
  {M.-C.}\ \bibnamefont {Tang}}, \bibinfo {author} {\bibfnamefont
  {M.}~\bibnamefont {Li}}, \bibinfo {author} {\bibfnamefont {Y.}~\bibnamefont
  {Shi}}, \bibinfo {author} {\bibfnamefont {W.}~\bibnamefont {Huang}}, \bibinfo
  {author} {\bibfnamefont {C.-W.}\ \bibnamefont {Qiu}},\ and\ \bibinfo {author}
  {\bibfnamefont {Q.}~\bibnamefont {Song}},\ }\bibfield  {title} {\bibinfo
  {title} {Arbitrarily polarized bound states in the continuum with twisted
  photonic crystal slabs},\ }\href {https://doi.org/10.1038/s41377-023-01090-w}
  {\bibfield  {journal} {\bibinfo  {journal} {Light: Science and Applications}\
  }\textbf {\bibinfo {volume} {12}},\ \bibinfo {pages} {1} (\bibinfo {year}
  {2023})}\BibitemShut {NoStop}%
\bibitem [{\citenamefont {Koshelev}\ and\ \citenamefont
  {Kivshar}(2019)}]{Koshelev2019a}%
  \BibitemOpen
  \bibfield  {author} {\bibinfo {author} {\bibfnamefont {K.}~\bibnamefont
  {Koshelev}}\ and\ \bibinfo {author} {\bibfnamefont {Y.}~\bibnamefont
  {Kivshar}},\ }\bibfield  {title} {\bibinfo {title} {Light trapping gets a
  boost},\ }\href {https://doi.org/10.1038/d41586-019-03143-w} {\bibfield
  {journal} {\bibinfo  {journal} {Nature}\ }\textbf {\bibinfo {volume} {574}},\
  \bibinfo {pages} {491} (\bibinfo {year} {2019})}\BibitemShut {NoStop}%
\bibitem [{\citenamefont {Wiersig}(2006)}]{Wiersig2006}%
  \BibitemOpen
  \bibfield  {author} {\bibinfo {author} {\bibfnamefont {J.}~\bibnamefont
  {Wiersig}},\ }\bibfield  {title} {\bibinfo {title} {Formation of long-lived,
  scarlike modes near avoided resonance crossings in optical microcavities},\
  }\href@noop {} {\bibfield  {journal} {\bibinfo  {journal} {Phys. Rev. Lett.}\
  }\textbf {\bibinfo {volume} {97}} (\bibinfo {year} {2006})}\BibitemShut
  {NoStop}%
\bibitem [{\citenamefont {Song}\ and\ \citenamefont {Cao}(2010)}]{Song2010}%
  \BibitemOpen
  \bibfield  {author} {\bibinfo {author} {\bibfnamefont {Q.~H.}\ \bibnamefont
  {Song}}\ and\ \bibinfo {author} {\bibfnamefont {H.}~\bibnamefont {Cao}},\
  }\bibfield  {title} {\bibinfo {title} {Improving optical confinement in
  nanostructures via external mode coupling},\ }\href
  {https://doi.org/10.1103/physrevlett.105.053902} {\bibfield  {journal}
  {\bibinfo  {journal} {Phys. Rev. Lett.}\ }\textbf {\bibinfo {volume} {105}},\
  \bibinfo {pages} {053902} (\bibinfo {year} {2010})}\BibitemShut {NoStop}%
\bibitem [{\citenamefont {Rybin}\ \emph {et~al.}(2017)\citenamefont {Rybin},
  \citenamefont {Koshelev}, \citenamefont {Sadrieva}, \citenamefont {Samusev},
  \citenamefont {Bogdanov}, \citenamefont {Limonov},\ and\ \citenamefont
  {Kivshar}}]{Rybin2017}%
  \BibitemOpen
  \bibfield  {author} {\bibinfo {author} {\bibfnamefont {M.}~\bibnamefont
  {Rybin}}, \bibinfo {author} {\bibfnamefont {K.}~\bibnamefont {Koshelev}},
  \bibinfo {author} {\bibfnamefont {Z.}~\bibnamefont {Sadrieva}}, \bibinfo
  {author} {\bibfnamefont {K.}~\bibnamefont {Samusev}}, \bibinfo {author}
  {\bibfnamefont {A.}~\bibnamefont {Bogdanov}}, \bibinfo {author}
  {\bibfnamefont {M.}~\bibnamefont {Limonov}},\ and\ \bibinfo {author}
  {\bibfnamefont {Y.}~\bibnamefont {Kivshar}},\ }\bibfield  {title} {\bibinfo
  {title} {High-{Q} {S}upercavity {M}odes in {S}ubwavelength {D}ielectric
  {R}esonators},\ }\href {https://doi.org/10.1103/physrevlett.119.243901}
  {\bibfield  {journal} {\bibinfo  {journal} {Phys. Rev. Lett.}\ }\textbf
  {\bibinfo {volume} {119}},\ \bibinfo {pages} {243901} (\bibinfo {year}
  {2017})}\BibitemShut {NoStop}%
\bibitem [{\citenamefont {Chen}\ \emph {et~al.}(2019)\citenamefont {Chen},
  \citenamefont {Chen},\ and\ \citenamefont {Liu}}]{Chen2019}%
  \BibitemOpen
  \bibfield  {author} {\bibinfo {author} {\bibfnamefont {W.}~\bibnamefont
  {Chen}}, \bibinfo {author} {\bibfnamefont {Y.}~\bibnamefont {Chen}},\ and\
  \bibinfo {author} {\bibfnamefont {W.}~\bibnamefont {Liu}},\ }\bibfield
  {title} {\bibinfo {title} {Multipolar conversion induced subwavelength high-q
  kerker supermodes with unidirectional radiations},\ }\href
  {https://doi.org/10.1002/lpor.201900067} {\bibfield  {journal} {\bibinfo
  {journal} {Laser and Photonics Reviews}\ }\textbf {\bibinfo {volume} {13}},\
  \bibinfo {pages} {1900067} (\bibinfo {year} {2019})}\BibitemShut {NoStop}%
\bibitem [{\citenamefont {Wang}\ \emph {et~al.}(2019)\citenamefont {Wang},
  \citenamefont {Zheng}, \citenamefont {Xiong}, \citenamefont {Qi},\ and\
  \citenamefont {Li}}]{Wang2019}%
  \BibitemOpen
  \bibfield  {author} {\bibinfo {author} {\bibfnamefont {W.}~\bibnamefont
  {Wang}}, \bibinfo {author} {\bibfnamefont {L.}~\bibnamefont {Zheng}},
  \bibinfo {author} {\bibfnamefont {L.}~\bibnamefont {Xiong}}, \bibinfo
  {author} {\bibfnamefont {J.}~\bibnamefont {Qi}},\ and\ \bibinfo {author}
  {\bibfnamefont {B.}~\bibnamefont {Li}},\ }\bibfield  {title} {\bibinfo
  {title} {High {Q}-factor multiple fano resonances for high-sensitivity
  sensing in all-dielectric metamaterials},\ }\href
  {https://doi.org/10.1364/osac.2.002818} {\bibfield  {journal} {\bibinfo
  {journal} {{OSA} Continuum}\ }\textbf {\bibinfo {volume} {2}},\ \bibinfo
  {pages} {2818} (\bibinfo {year} {2019})}\BibitemShut {NoStop}%
\bibitem [{\citenamefont {Odit}\ \emph {et~al.}(2020)\citenamefont {Odit},
  \citenamefont {Koshelev}, \citenamefont {Gladyshev}, \citenamefont
  {Ladutenko}, \citenamefont {Kivshar},\ and\ \citenamefont
  {Bogdanov}}]{Odit2020}%
  \BibitemOpen
  \bibfield  {author} {\bibinfo {author} {\bibfnamefont {M.}~\bibnamefont
  {Odit}}, \bibinfo {author} {\bibfnamefont {K.}~\bibnamefont {Koshelev}},
  \bibinfo {author} {\bibfnamefont {S.}~\bibnamefont {Gladyshev}}, \bibinfo
  {author} {\bibfnamefont {K.}~\bibnamefont {Ladutenko}}, \bibinfo {author}
  {\bibfnamefont {Y.}~\bibnamefont {Kivshar}},\ and\ \bibinfo {author}
  {\bibfnamefont {A.}~\bibnamefont {Bogdanov}},\ }\bibfield  {title} {\bibinfo
  {title} {Observation of supercavity modes in subwavelength dielectric
  resonators},\ }\href {https://doi.org/10.1002/adma.202003804} {\bibfield
  {journal} {\bibinfo  {journal} {Advanced Materials}\ ,\ \bibinfo {pages}
  {2003804}} (\bibinfo {year} {2020})}\BibitemShut {NoStop}%
\bibitem [{\citenamefont {Volkovskaya}\ \emph {et~al.}(2020)\citenamefont
  {Volkovskaya}, \citenamefont {Xu}, \citenamefont {Huang}, \citenamefont
  {Smirnov}, \citenamefont {Miroshnichenko},\ and\ \citenamefont
  {Smirnova}}]{Volkovskaya2020}%
  \BibitemOpen
  \bibfield  {author} {\bibinfo {author} {\bibfnamefont {I.}~\bibnamefont
  {Volkovskaya}}, \bibinfo {author} {\bibfnamefont {L.}~\bibnamefont {Xu}},
  \bibinfo {author} {\bibfnamefont {L.}~\bibnamefont {Huang}}, \bibinfo
  {author} {\bibfnamefont {A.~I.}\ \bibnamefont {Smirnov}}, \bibinfo {author}
  {\bibfnamefont {A.~E.}\ \bibnamefont {Miroshnichenko}},\ and\ \bibinfo
  {author} {\bibfnamefont {D.}~\bibnamefont {Smirnova}},\ }\bibfield  {title}
  {\bibinfo {title} {Multipolar second-harmonic generation from high-q
  quasi-{BIC} states in subwavelength resonators},\ }\href
  {https://doi.org/10.1515/nanoph-2020-0156} {\bibfield  {journal} {\bibinfo
  {journal} {Nanophotonics}\ }\textbf {\bibinfo {volume} {9}},\ \bibinfo
  {pages} {3953} (\bibinfo {year} {2020})}\BibitemShut {NoStop}%
\bibitem [{\citenamefont {Huang}\ \emph {et~al.}(2021)\citenamefont {Huang},
  \citenamefont {Xu}, \citenamefont {Rahmani}, \citenamefont {Neshev},\ and\
  \citenamefont {Miroshnichenko}}]{Huang2021}%
  \BibitemOpen
  \bibfield  {author} {\bibinfo {author} {\bibfnamefont {L.}~\bibnamefont
  {Huang}}, \bibinfo {author} {\bibfnamefont {L.}~\bibnamefont {Xu}}, \bibinfo
  {author} {\bibfnamefont {M.}~\bibnamefont {Rahmani}}, \bibinfo {author}
  {\bibfnamefont {D.}~\bibnamefont {Neshev}},\ and\ \bibinfo {author}
  {\bibfnamefont {A.}~\bibnamefont {Miroshnichenko}},\ }\bibfield  {title}
  {\bibinfo {title} {Pushing the limit of high-{Q} mode of a single dielectric
  nanocavity},\ }\href {https://doi.org/10.1117/1.ap.3.1.016004} {\bibfield
  {journal} {\bibinfo  {journal} {Advanced Photonics}\ }\textbf {\bibinfo
  {volume} {3}},\ \bibinfo {pages} {016004} (\bibinfo {year}
  {2021})}\BibitemShut {NoStop}%
\bibitem [{\citenamefont {Bulgakov}\ and\ \citenamefont
  {Sadreev}(2014)}]{Bulgakov2014}%
  \BibitemOpen
  \bibfield  {author} {\bibinfo {author} {\bibfnamefont {E.~N.}\ \bibnamefont
  {Bulgakov}}\ and\ \bibinfo {author} {\bibfnamefont {A.~F.}\ \bibnamefont
  {Sadreev}},\ }\bibfield  {title} {\bibinfo {title} {Bloch bound states in the
  radiation continuum in a periodic array of dielectric rods},\ }\href@noop {}
  {\bibfield  {journal} {\bibinfo  {journal} {Phys. Rev. A}\ }\textbf {\bibinfo
  {volume} {90}},\ \bibinfo {pages} {053801} (\bibinfo {year}
  {2014})}\BibitemShut {NoStop}%
\bibitem [{\citenamefont {{Zhen Hu}}\ and\ \citenamefont {{Ya Yan
  Lu}}(2015)}]{Hu&Lu2015}%
  \BibitemOpen
  \bibfield  {author} {\bibinfo {author} {\bibnamefont {{Zhen Hu}}}\ and\
  \bibinfo {author} {\bibnamefont {{Ya Yan Lu}}},\ }\bibfield  {title}
  {\bibinfo {title} {Standing waves on two-dimensional periodic dielectric
  waveguides},\ }\href@noop {} {\bibfield  {journal} {\bibinfo  {journal} {J.
  Optics}\ }\textbf {\bibinfo {volume} {17}},\ \bibinfo {pages} {065601}
  (\bibinfo {year} {2015})}\BibitemShut {NoStop}%
\bibitem [{\citenamefont {Vincent}\ and\ \citenamefont
  {Nevi{`e}re}(1979)}]{Vincent1979}%
  \BibitemOpen
  \bibfield  {author} {\bibinfo {author} {\bibfnamefont {P.}~\bibnamefont
  {Vincent}}\ and\ \bibinfo {author} {\bibfnamefont {M.}~\bibnamefont
  {Nevi{`e}re}},\ }\bibfield  {title} {\bibinfo {title} {Corrugated dielectric
  waveguides: A numerical study of the second-order stop bands},\ }\href
  {https://doi.org/10.1007/bf00895008} {\bibfield  {journal} {\bibinfo
  {journal} {Appl. Phys.}\ }\textbf {\bibinfo {volume} {20}},\ \bibinfo {pages}
  {345} (\bibinfo {year} {1979})}\BibitemShut {NoStop}%
\bibitem [{\citenamefont {{Yang Yi}}\ \emph {et~al.}(2014)\citenamefont {{Yang
  Yi}}, \citenamefont {{Peng Chao}}, \citenamefont {{Liang Yong}},
  \citenamefont {{Li Zhengbin}},\ and\ \citenamefont {Noda}}]{Yang2014}%
  \BibitemOpen
  \bibfield  {author} {\bibinfo {author} {\bibnamefont {{Yang Yi}}}, \bibinfo
  {author} {\bibnamefont {{Peng Chao}}}, \bibinfo {author} {\bibnamefont
  {{Liang Yong}}}, \bibinfo {author} {\bibnamefont {{Li Zhengbin}}},\ and\
  \bibinfo {author} {\bibfnamefont {S.}~\bibnamefont {Noda}},\ }\bibfield
  {title} {\bibinfo {title} {Analytical perspective for bound states in the
  continuum in photonic crystal slabs},\ }\href
  {https://doi.org/10.1103/physrevlett.113.037401} {\bibfield  {journal}
  {\bibinfo  {journal} {Physical Review Letters}\ }\textbf {\bibinfo {volume}
  {113}},\ \bibinfo {pages} {037401} (\bibinfo {year} {2014})}\BibitemShut
  {NoStop}%
\bibitem [{\citenamefont {Bykov}\ and\ \citenamefont
  {Doskolovich}(2015)}]{Bykov2015}%
  \BibitemOpen
  \bibfield  {author} {\bibinfo {author} {\bibfnamefont {D.}~\bibnamefont
  {Bykov}}\ and\ \bibinfo {author} {\bibfnamefont {L.}~\bibnamefont
  {Doskolovich}},\ }\bibfield  {title} {\bibinfo {title} {{$\omega-kx$ Fano}
  line shape in photonic crystal slabs},\ }\href
  {https://doi.org/10.1103/physreva.92.013845} {\bibfield  {journal} {\bibinfo
  {journal} {Phys. Rev. A}\ }\textbf {\bibinfo {volume} {92}},\ \bibinfo
  {pages} {013845} (\bibinfo {year} {2015})}\BibitemShut {NoStop}%
\bibitem [{\citenamefont {Gao}\ \emph {et~al.}(2016)\citenamefont {Gao},
  \citenamefont {Hsu}, \citenamefont {Zhen}, \citenamefont {Lin}, \citenamefont
  {Joannopoulos}, \citenamefont {Solja{\v{c}}i{\'{c}}},\ and\ \citenamefont
  {Chen}}]{Gao2016}%
  \BibitemOpen
  \bibfield  {author} {\bibinfo {author} {\bibfnamefont {X.}~\bibnamefont
  {Gao}}, \bibinfo {author} {\bibfnamefont {C.~W.}\ \bibnamefont {Hsu}},
  \bibinfo {author} {\bibfnamefont {B.}~\bibnamefont {Zhen}}, \bibinfo {author}
  {\bibfnamefont {X.}~\bibnamefont {Lin}}, \bibinfo {author} {\bibfnamefont
  {J.~D.}\ \bibnamefont {Joannopoulos}}, \bibinfo {author} {\bibfnamefont
  {M.}~\bibnamefont {Solja{\v{c}}i{\'{c}}}},\ and\ \bibinfo {author}
  {\bibfnamefont {H.}~\bibnamefont {Chen}},\ }\bibfield  {title} {\bibinfo
  {title} {Formation mechanism of guided resonances and bound states in the
  continuum in photonic crystal slabs},\ }\href
  {https://doi.org/10.1038/srep31908} {\bibfield  {journal} {\bibinfo
  {journal} {Scientific Reports}\ }\textbf {\bibinfo {volume} {6}},\ \bibinfo
  {pages} {1} (\bibinfo {year} {2016})}\BibitemShut {NoStop}%
\bibitem [{\citenamefont {Ni}\ \emph {et~al.}(2016)\citenamefont {Ni},
  \citenamefont {Wang}, \citenamefont {Peng},\ and\ \citenamefont
  {Li}}]{Ni2016}%
  \BibitemOpen
  \bibfield  {author} {\bibinfo {author} {\bibfnamefont {L.}~\bibnamefont
  {Ni}}, \bibinfo {author} {\bibfnamefont {Z.}~\bibnamefont {Wang}}, \bibinfo
  {author} {\bibfnamefont {C.}~\bibnamefont {Peng}},\ and\ \bibinfo {author}
  {\bibfnamefont {Z.}~\bibnamefont {Li}},\ }\bibfield  {title} {\bibinfo
  {title} {Tunable optical bound states in the continuum beyond in-plane
  symmetry protection},\ }\bibfield  {journal} {\bibinfo  {journal} {Physical
  Review B}\ }\textbf {\bibinfo {volume} {94}},\ \href
  {https://doi.org/10.1103/physrevb.94.245148} {10.1103/physrevb.94.245148}
  (\bibinfo {year} {2016})\BibitemShut {NoStop}%
\bibitem [{\citenamefont {Friedrich}\ and\ \citenamefont
  {Wintgen}(1985)}]{Friedrich1985}%
  \BibitemOpen
  \bibfield  {author} {\bibinfo {author} {\bibfnamefont {H.}~\bibnamefont
  {Friedrich}}\ and\ \bibinfo {author} {\bibfnamefont {D.}~\bibnamefont
  {Wintgen}},\ }\bibfield  {title} {\bibinfo {title} {Interfering resonances
  and bound states in the continuum},\ }\href
  {https://doi.org/10.1103/physreva.32.3231} {\bibfield  {journal} {\bibinfo
  {journal} {Phys. Rev. A}\ }\textbf {\bibinfo {volume} {32}},\ \bibinfo
  {pages} {3231} (\bibinfo {year} {1985})}\BibitemShut {NoStop}%
\bibitem [{\citenamefont {Volya}\ and\ \citenamefont
  {Zelevinsky}(2003)}]{Volya}%
  \BibitemOpen
  \bibfield  {author} {\bibinfo {author} {\bibfnamefont {A.}~\bibnamefont
  {Volya}}\ and\ \bibinfo {author} {\bibfnamefont {V.}~\bibnamefont
  {Zelevinsky}},\ }\bibfield  {title} {\bibinfo {title} {Non-hermitian
  effective hamiltonian and continuum shell model},\ }\href
  {https://doi.org/10.1103/physrevc.67.054322} {\bibfield  {journal} {\bibinfo
  {journal} {Phys. Rev. C}\ }\textbf {\bibinfo {volume} {67}},\ \bibinfo
  {pages} {054322} (\bibinfo {year} {2003})}\BibitemShut {NoStop}%
\bibitem [{\citenamefont {Sadreev}(2021)}]{Sadreev2021}%
  \BibitemOpen
  \bibfield  {author} {\bibinfo {author} {\bibfnamefont {A.}~\bibnamefont
  {Sadreev}},\ }\bibfield  {title} {\bibinfo {title} {Interference traps waves
  in an open system: bound states in the continuum},\ }\href
  {https://doi.org/10.1088/1361-6633/abefb9} {\bibfield  {journal} {\bibinfo
  {journal} {Rep. Progr. Phys.}\ }\textbf {\bibinfo {volume} {84}},\ \bibinfo
  {pages} {055901} (\bibinfo {year} {2021})}\BibitemShut {NoStop}%
\bibitem [{\citenamefont {Kikkawa}\ \emph {et~al.}(2019)\citenamefont
  {Kikkawa}, \citenamefont {Nishida},\ and\ \citenamefont
  {Kadoya}}]{Kikkawa2019}%
  \BibitemOpen
  \bibfield  {author} {\bibinfo {author} {\bibfnamefont {R.}~\bibnamefont
  {Kikkawa}}, \bibinfo {author} {\bibfnamefont {M.}~\bibnamefont {Nishida}},\
  and\ \bibinfo {author} {\bibfnamefont {Y.}~\bibnamefont {Kadoya}},\
  }\bibfield  {title} {\bibinfo {title} {Polarization-based branch selection of
  bound states in the continuum in dielectric waveguide modes anti-crossed by a
  metal grating},\ }\href {https://doi.org/10.1088/1367-2630/ab4f54} {\bibfield
   {journal} {\bibinfo  {journal} {New Journal of Physics}\ }\textbf {\bibinfo
  {volume} {21}},\ \bibinfo {pages} {113020} (\bibinfo {year}
  {2019})}\BibitemShut {NoStop}%
\bibitem [{\citenamefont {{S. Gladyshev}}\ \emph {et~al.}(2022)\citenamefont
  {{S. Gladyshev}}, \citenamefont {{A. Shalev}}, \citenamefont {{K.
  Ladutenko}},\ and\ \citenamefont {{A. Bogdanov}}}]{Gladyshev2022}%
  \BibitemOpen
  \bibfield  {author} {\bibinfo {author} {\bibnamefont {{S. Gladyshev}}},
  \bibinfo {author} {\bibnamefont {{A. Shalev}}}, \bibinfo {author}
  {\bibfnamefont {K.}~\bibnamefont {{K. Ladutenko}}},\ and\ \bibinfo {author}
  {\bibnamefont {{A. Bogdanov}}},\ }\bibfield  {title} {\bibinfo {title} {Bound
  states in the continuum in multipolar lattices},\ }\href@noop {} {\bibfield
  {journal} {\bibinfo  {journal} {Phys. Rev. B}\ }\textbf {\bibinfo {volume}
  {105}},\ \bibinfo {pages} {L241301} (\bibinfo {year} {2022})}\BibitemShut
  {NoStop}%
\bibitem [{\citenamefont {Yasumoto}\ and\ \citenamefont
  {Jia}(2006)}]{Yasumoto}%
  \BibitemOpen
  \bibfield  {author} {\bibinfo {author} {\bibfnamefont {K.}~\bibnamefont
  {Yasumoto}}\ and\ \bibinfo {author} {\bibfnamefont {H.}~\bibnamefont {Jia}},\
  }\bibfield  {title} {\bibinfo {title} {Modeling of photonic crystals by
  multilayered periodic arrays of circular cylinders},\ }in\ \href@noop {}
  {\emph {\bibinfo {booktitle} {Electromagnetic Theory and Applications for
  Photonic Crystals}}},\ \bibinfo {editor} {edited by\ \bibinfo {editor}
  {\bibfnamefont {K.}~\bibnamefont {Yasumoto}}}\ (\bibinfo  {publisher} {MIT
  Press},\ \bibinfo {address} {Cambridge, MA},\ \bibinfo {year} {2006})\ p.\
  \bibinfo {pages} {135 Eq.(3.50)}\BibitemShut {NoStop}%
\bibitem [{\citenamefont {Johnson}\ \emph {et~al.}(2001)\citenamefont
  {Johnson}, \citenamefont {Fan}, \citenamefont {Mekis},\ and\ \citenamefont
  {Joannopoulos}}]{Johnson2001}%
  \BibitemOpen
  \bibfield  {author} {\bibinfo {author} {\bibfnamefont {S.~G.}\ \bibnamefont
  {Johnson}}, \bibinfo {author} {\bibfnamefont {S.}~\bibnamefont {Fan}},
  \bibinfo {author} {\bibfnamefont {A.}~\bibnamefont {Mekis}},\ and\ \bibinfo
  {author} {\bibfnamefont {J.~D.}\ \bibnamefont {Joannopoulos}},\ }\bibfield
  {title} {\bibinfo {title} {Multipole-cancellation mechanism for high-q
  cavities in the absence of a complete photonic band gap},\ }\href
  {https://doi.org/10.1063/1.1375838} {\bibfield  {journal} {\bibinfo
  {journal} {Appl. Phys. Lett.}\ }\textbf {\bibinfo {volume} {78}},\ \bibinfo
  {pages} {3388} (\bibinfo {year} {2001})}\BibitemShut {NoStop}%
\bibitem [{\citenamefont {Blaustein}\ \emph {et~al.}(2007)\citenamefont
  {Blaustein}, \citenamefont {Gozman}, \citenamefont {Samoylova}, \citenamefont
  {Polishchuk},\ and\ \citenamefont {Burin}}]{Blaustein2007}%
  \BibitemOpen
  \bibfield  {author} {\bibinfo {author} {\bibfnamefont {G.}~\bibnamefont
  {Blaustein}}, \bibinfo {author} {\bibfnamefont {M.}~\bibnamefont {Gozman}},
  \bibinfo {author} {\bibfnamefont {O.}~\bibnamefont {Samoylova}}, \bibinfo
  {author} {\bibfnamefont {I.}~\bibnamefont {Polishchuk}},\ and\ \bibinfo
  {author} {\bibfnamefont {A.}~\bibnamefont {Burin}},\ }\bibfield  {title}
  {\bibinfo {title} {Guiding optical modes in chains of dielectric particles},\
  }\href {https://doi.org/10.1364/oe.15.017380} {\bibfield  {journal} {\bibinfo
   {journal} {Optics Express}\ }\textbf {\bibinfo {volume} {15}},\ \bibinfo
  {pages} {17380} (\bibinfo {year} {2007})}\BibitemShut {NoStop}%
\bibitem [{\citenamefont {Asenjo-Garcia}\ \emph {et~al.}(2017)\citenamefont
  {Asenjo-Garcia}, \citenamefont {Moreno-Cardoner}, \citenamefont {Albrecht},
  \citenamefont {Kimble},\ and\ \citenamefont {Chang}}]{Asenjo2017}%
  \BibitemOpen
  \bibfield  {author} {\bibinfo {author} {\bibfnamefont {A.}~\bibnamefont
  {Asenjo-Garcia}}, \bibinfo {author} {\bibfnamefont {M.}~\bibnamefont
  {Moreno-Cardoner}}, \bibinfo {author} {\bibfnamefont {A.}~\bibnamefont
  {Albrecht}}, \bibinfo {author} {\bibfnamefont {H.}~\bibnamefont {Kimble}},\
  and\ \bibinfo {author} {\bibfnamefont {D.}~\bibnamefont {Chang}},\ }\bibfield
   {title} {\bibinfo {title} {Exponential improvement in photon storage
  fidelities using subradiance and {\textquotedblleft}selective
  radiance{\textquotedblright} in atomic arrays},\ }\bibfield  {journal}
  {\bibinfo  {journal} {Phys. Rev. X}\ }\textbf {\bibinfo {volume} {7}},\ \href
  {https://doi.org/10.1103/physrevx.7.031024} {10.1103/physrevx.7.031024}
  (\bibinfo {year} {2017})\BibitemShut {NoStop}%
\bibitem [{\citenamefont {Bulgakov}\ and\ \citenamefont
  {Sadreev}(2019)}]{Bulgakov2019a}%
  \BibitemOpen
  \bibfield  {author} {\bibinfo {author} {\bibfnamefont {E.}~\bibnamefont
  {Bulgakov}}\ and\ \bibinfo {author} {\bibfnamefont {A.}~\bibnamefont
  {Sadreev}},\ }\bibfield  {title} {\bibinfo {title} {High-{Q} resonant modes
  in a finite array of dielectric particles},\ }\href
  {https://doi.org/10.1103/physreva.99.033851} {\bibfield  {journal} {\bibinfo
  {journal} {Phys. Rev. A}\ }\textbf {\bibinfo {volume} {99}},\ \bibinfo
  {pages} {033851} (\bibinfo {year} {2019})}\BibitemShut {NoStop}%
\bibitem [{\citenamefont {Zhang}\ \emph {et~al.}(2011)\citenamefont {Zhang},
  \citenamefont {Bulu}, \citenamefont {Tam}, \citenamefont {Levitt},
  \citenamefont {Shah}, \citenamefont {Botto},\ and\ \citenamefont
  {Loncar}}]{Zhang2011}%
  \BibitemOpen
  \bibfield  {author} {\bibinfo {author} {\bibfnamefont {Y.}~\bibnamefont
  {Zhang}}, \bibinfo {author} {\bibfnamefont {I.}~\bibnamefont {Bulu}},
  \bibinfo {author} {\bibfnamefont {W.-M.}\ \bibnamefont {Tam}}, \bibinfo
  {author} {\bibfnamefont {B.}~\bibnamefont {Levitt}}, \bibinfo {author}
  {\bibfnamefont {J.}~\bibnamefont {Shah}}, \bibinfo {author} {\bibfnamefont
  {T.}~\bibnamefont {Botto}},\ and\ \bibinfo {author} {\bibfnamefont
  {M.}~\bibnamefont {Loncar}},\ }\bibfield  {title} {\bibinfo {title} {High-q/v
  air-mode photonic crystal cavities at microwave frequencies},\ }\href
  {https://doi.org/10.1364/oe.19.009371} {\bibfield  {journal} {\bibinfo
  {journal} {Optics Express}\ }\textbf {\bibinfo {volume} {19}},\ \bibinfo
  {pages} {9371} (\bibinfo {year} {2011})}\BibitemShut {NoStop}%
\bibitem [{\citenamefont {Kornovan}\ \emph {et~al.}(2021)\citenamefont
  {Kornovan}, \citenamefont {Savelev}, \citenamefont {Kivshar},\ and\
  \citenamefont {Petrov}}]{Kornovan2021}%
  \BibitemOpen
  \bibfield  {author} {\bibinfo {author} {\bibfnamefont {D.}~\bibnamefont
  {Kornovan}}, \bibinfo {author} {\bibfnamefont {R.}~\bibnamefont {Savelev}},
  \bibinfo {author} {\bibfnamefont {Y.}~\bibnamefont {Kivshar}},\ and\ \bibinfo
  {author} {\bibfnamefont {M.}~\bibnamefont {Petrov}},\ }\bibfield  {title}
  {\bibinfo {title} {High-{Q} localized states in finite arrays of
  subwavelength resonators},\ }\href
  {https://doi.org/10.1021/acsphotonics.1c01262} {\bibfield  {journal}
  {\bibinfo  {journal} {{ACS} Photonics}\ }\textbf {\bibinfo {volume} {8}},\
  \bibinfo {pages} {3627} (\bibinfo {year} {2021})}\BibitemShut {NoStop}%
\bibitem [{\citenamefont {Kikkawa}\ \emph {et~al.}(2020)\citenamefont
  {Kikkawa}, \citenamefont {Nishida},\ and\ \citenamefont
  {Kadoya}}]{Kikkawa2020}%
  \BibitemOpen
  \bibfield  {author} {\bibinfo {author} {\bibfnamefont {R.}~\bibnamefont
  {Kikkawa}}, \bibinfo {author} {\bibfnamefont {M.}~\bibnamefont {Nishida}},\
  and\ \bibinfo {author} {\bibfnamefont {Y.}~\bibnamefont {Kadoya}},\
  }\bibfield  {title} {\bibinfo {title} {Bound states in the continuum and
  exceptional points in dielectric waveguide equipped with a metal grating},\
  }\href {https://doi.org/10.1088/1367-2630/ab97e9} {\bibfield  {journal}
  {\bibinfo  {journal} {New Journal of Physics}\ }\textbf {\bibinfo {volume}
  {22}},\ \bibinfo {pages} {073029} (\bibinfo {year} {2020})}\BibitemShut
  {NoStop}%
\end{thebibliography}
%

\end{document}